\documentclass[onecolumn,aps,pra,reprint,superscriptaddress,longbibliography]{revtex4-1}
\usepackage{amsmath}
\usepackage{graphicx}
\usepackage[lofdepth,lotdepth, caption=false]{subfig}
\usepackage{verbatim}
\usepackage{color}
\usepackage{dcolumn}
\usepackage{bm}
\usepackage{miller}
\usepackage{color}

\usepackage{setspace}

\usepackage{xr-hyper}
\usepackage{hyperref}

\usepackage[english]{babel}

\newcommand{\beginsupplement}{%
        \setcounter{table}{0}
        \renewcommand{\thetable}{S\arabic{table}}%
        \setcounter{figure}{0}
        \renewcommand{\thefigure}{S\arabic{figure}}%
     }

\begin{document}

\title{Antiferromagnetic-ferromagnetic homostructures with Dirac magnons in van der Waals  magnet CrI$_3$}

\author{John A.~Schneeloch}
\affiliation{Department of Physics, University of Virginia, Charlottesville,
Virginia 22904, USA}
%\email{jas9db@virginia.edu}

\author{Luke Daemen}
\affiliation{Neutron Scattering Division, Oak Ridge National Laboratory, Oak Ridge, Tennessee 37831, USA}

\author{Despina Louca}
\thanks{Corresponding author}
\email{louca@virginia.edu}

\begin{abstract}
Van der Waals (vdW) Dirac magnon system CrI$_{3}$, a potential host of topological edge magnons, orders ferromagnetically (FM) (T$_{c}$=61 K) in the bulk, but antiferromagnetic (AFM) order has been observed in nanometer thick flakes, attributed to monoclinic (M) type stacking.
We report neutron scattering measurements on a powder sample where the usual transition to the rhombohedral (R) phase was inhibited for a majority of the structure. Elastic measurements (and the opening of a hysteresis in magnetization data on a pressed pellet) showed that an AFM transition is clearly present below $\sim$50 K, coexisting with the R-type FM order.
Inelastic measurements showed a decrease in magnon energy compared to the R phase, consistent with a smaller interlayer magnetic coupling in M-type stacking. 
A gap remains at the Dirac point, suggesting that the same nontrivial magnon topology reported for the R phase may be present in the M phase as well.

\end{abstract}

\maketitle

\pagestyle{plain}

%\section{Introduction}
The exfoliation of single atomic layers from bulk vdW crystals has revolutionized device concepts based on heterostructures \cite{novoselovElectricFieldEffect2004}. The observation of superconductivity in twisted graphene \cite{caoUnconventionalSuperconductivityMagicangle2018} is an example of what single layer manipulation can do. Giant magnetoresistance (GMR) preceded the concept of heterostructure layering, built on alternating layers of FM with AFM order \cite{baibichGiantMagnetoresistance0011988,binaschEnhancedMagnetoresistanceLayered1989}. More recently, 
multiple stacking possibilities in vdW crystals have been shown to lead to markedly different behaviors. For instance, transitions from non-trivial to topological band structures have been observed in the Weyl semimetal MoTe$_{2}$ from the $1T^{\prime}$ (monoclinic) to $T_d$ (orthorhombic) phases \cite{sunPredictionWeylSemimetal2015,dengExperimentalObservationTopological2016} or from a weak to a strong topological insulator in Bi$_{4}$I$_{4}$ \cite{noguchiWeakTopologicalInsulator2019,huangRoomtemperatureTopologicalPhase2021}. The magnetic behavior may change with the stacking as well, such as the roughly tenfold increase in the interlayer magnetic coupling reported for CrCl$_{3}$ when the layer stacking present at high temperatures is preserved at low temperatures \cite{kleinEnhancementInterlayerExchange2019,serriEnhancementMagneticCoupling2020}. 

\begin{figure}[h]
\begin{center}
\includegraphics[width=8.6cm]
{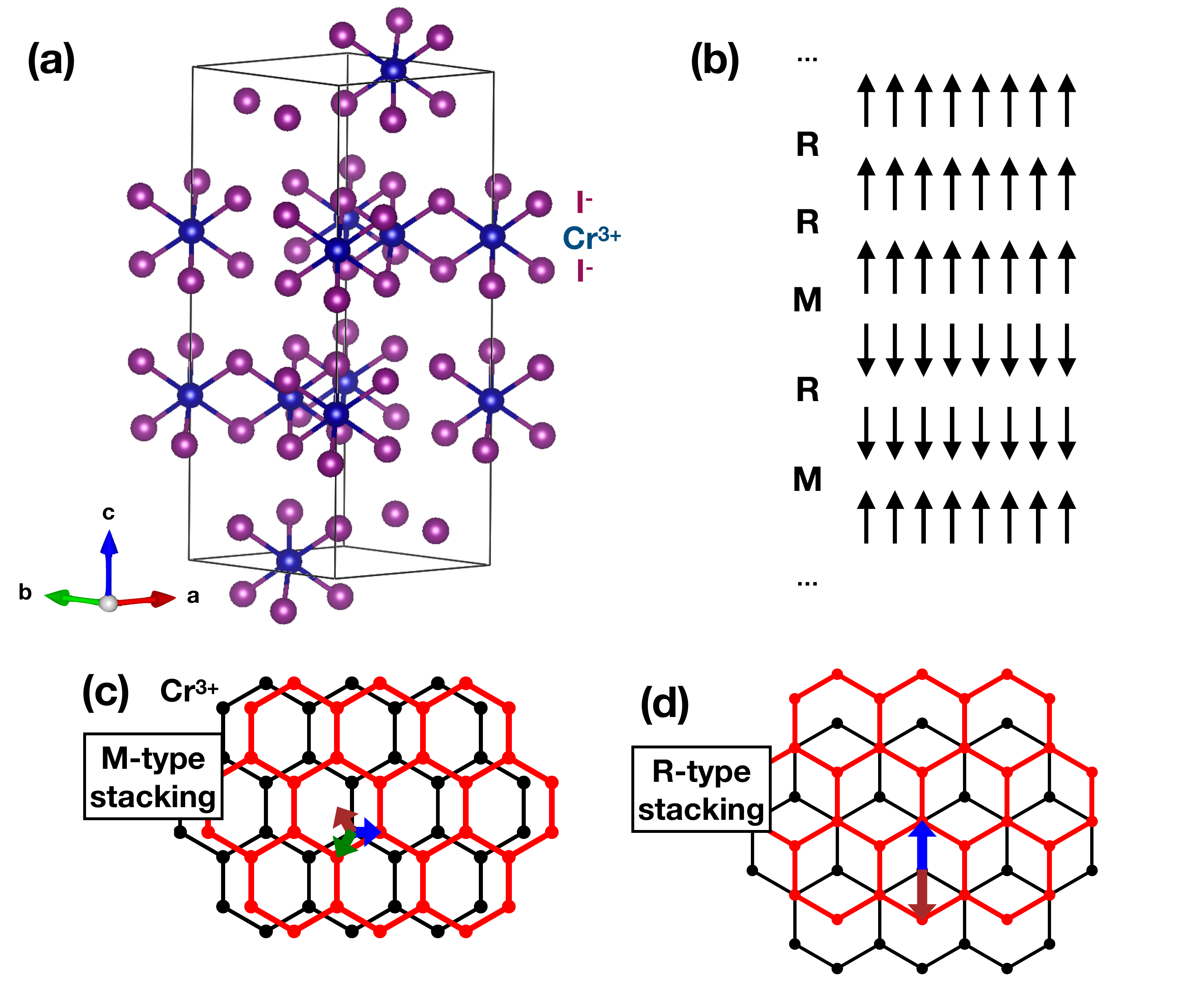}
\end{center}
\caption{(a) Crystal structure of the $R\bar{3}$ phase of CrI$_{3}$. (b) A schematic showing how the spin direction would change layer-by-layer in M and R type stacking, with spin flips accompanying M-type stacking. (c,d) Illustration of (c) M-type and (d) R-type layer stacking. The honeycomb lattice represent the placement of the Cr$^{3+}$ ions, with the red lattice above the black showing one possible stacking option; the displacements for the full set of stacking options are shown as arrows.}
\label{fig:1}
\end{figure}

CrI$_{3}$ consists of layers of honeycomb lattices of Cr$^{3+}$ ions with $S=3/2$ spins sandwiched between two triangular lattices of I$^{-}$ ions. The I$^{-}$ ions of one layer sit in the middle of the triangles of the neighboring I$^{-}$ lattice (Fig.\ \ref{fig:1}(a)). Bulk crystals become FM below $T_C = 61$ K, with the spins oriented out-of-plane \cite{mcguireCouplingCrystalStructure2015}. In thin flakes, on the other hand, the spin alignment is AFM with the spin direction (pointing out-of-plane) alternating layer-by-layer \cite{huangLayerdependentFerromagnetismVan2017}, as deduced from techniques such as the magneto-optical Kerr effect \cite{huangLayerdependentFerromagnetismVan2017,wangVeryLargeTunneling2018}, magnetic circular dichroism \cite{jiangElectricfieldSwitchingTwodimensional2018,songSwitching2DMagnetic2019,liPressurecontrolledInterlayerMagnetism2019}, magnetic force microscopy \cite{niuCoexistenceMagneticOrders2020}, scanning magnetometry \cite{thielProbingMagnetism2D2019}, and tunneling magnetoresistance \cite{songGiantTunnelingMagnetoresistance2018,kleinProbingMagnetism2D2018,wangVeryLargeTunneling2018,kimOneMillionPercent2018}.

The AFM order in flakes of CrI$_{3}$ arises from monoclinic $C2/m$ (M-type) stacking that is present because of the arrested transition to the rhombohedral $R\bar{3}$ (R-type) stacking \cite{ubrigLowtemperatureMonoclinicLayer2019,liPressurecontrolledInterlayerMagnetism2019,wangVeryLargeTunneling2018,sivadasStackingdependentMagnetismBilayer2018,sorianoInterplayInterlayerExchange2019,jiangStackingTunableInterlayer2019,jangMicroscopicUnderstandingMagnetic2019,sunGiantNonreciprocalSecondharmonic2019}. 
There are two sets of symmetry-equivalent stacking possibilities, with three M-type and two R-type stacking options (Fig.\ \ref{fig:1}(c,d).)
In principle, the M-type stacking disappears in bulk CrI$_{3}$ on cooling during the layer-sliding transition from the M-phase to the R-phase below $\sim$180 K \cite{mcguireCouplingCrystalStructure2015,djurdjic-mijinLatticeDynamicsPhase2018}.

However, even in single crystals, the M$\rightarrow$R transition may occur over a  broad temperature range or be inhibited entirely (as can be seen in our single-crystal x-ray diffraction measurements in Fig.\ \ref{fig:SingleCrystalXRD}), and the process may proceed differently in subsequent thermal cycles \cite{mcguireCouplingCrystalStructure2015}. The thickness of crystals is also an important factor in inhibiting the transition, as shown from split vibrational modes, observed in the Raman spectroscopy of a thin flake, that fail to merge (as expected for $R\bar{3}$) down to at least 10 K \cite{guoStructuralMonoclinicityIts2021}. The dependence on thickness of layer-sliding transitions has also been observed in MoTe$_{2}$, where the transition temperature range broadens (or is inhibited entirely) for crystals with a thickness below $\sim$120 nm \cite{caoBarkhausenEffectFirst2018, heDimensionalitydrivenOrthorhombicMoTe22018}.) Surface layers of CrI$_{3}$ crystals have also been reported to exhibit AFM ordering, presumably from M-type stacking \cite{liMagneticfieldinducedQuantumPhase2020,niuCoexistenceMagneticOrders2020,suCurrentinducedCrI3Surface2020}. At the same time, several bulk measurements hint at the presence of magnetic ordering beyond the reported ferromagnetism, such as the existence of anomalies near $\sim$50 K in magnetic susceptibility \cite{liuAnisotropicMagnetocaloricEffect2018,wangVeryLargeTunneling2018,mcguireCouplingCrystalStructure2015,arnethUniaxialPressureEffects2022}, and a second component evident in muon spin resonance ($\mu$SR) measurements \cite{meseguer-sanchezCoexistenceStructuralMagnetic2021}. 

CrI$_{3}$ is also a candidate material for observing topological magnons \cite{mcclartyTopologicalMagnonsReview2022}. CrI$_{3}$ has been probed via inelastic neutron scattering in several recent studies \cite{chenTopologicalSpinExcitations2018, chenMagneticAnisotropyFerromagnetic2020, chenMagneticFieldEffect2021, chenMasslessDiracMagnons2021}, in which the spin waves were described in terms of a dispersion reminiscent of the electronic band structure of graphene, but with the Dzyaloshinskii-Moriya (DM) interaction reportedly opening a gap of 2.8 meV at the Dirac points \cite{chenMagneticFieldEffect2021}.
With neutron scattering, we elucidate the dual magnetic nature of CrI${_3}$ by providing direct evidence of M-type stacking with AFM order that alternates with R-type stacking with FM order in bulk samples. Elastic neutron scattering measurements on ground CrI$_{3}$ powder show magnetic elastic intensity that is consistent with a model where the spin direction flips across M-type interlayer boundaries. Thus, control of the M-to-R layering can provide a homostructure with AFM-to-FM order. The AFM ordering vanishes above $\sim$50-55 K, while the FM ordering persists to $\sim$60 K. From inelastic neutron scattering, a 
$\lesssim$1 meV decrease in energy relative to the reported single-crystal dispersion is observed. 
A gap is present at the Dirac node, suggesting its presence (and the possibility of topological magnons) in the $M$ phase as well as the $R$ phase.

\section*{Elastic neutron scattering}
\begin{figure}[h]
\begin{center}
\includegraphics[width=8.6cm]
{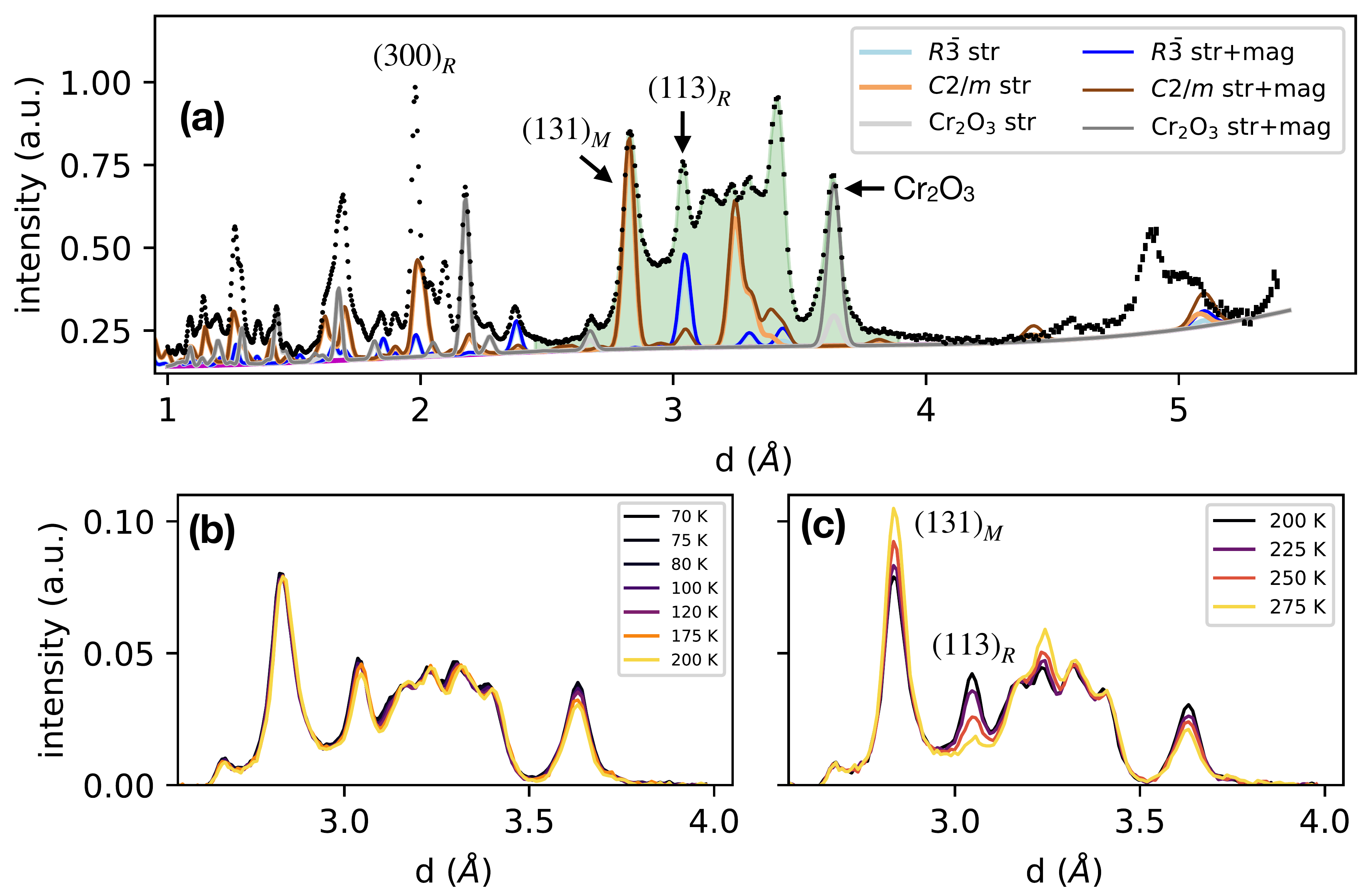}
\end{center}
\caption{
(a) Data at 5 K (black points), along with curves of simulated intensity for the $R\bar{3}$ and $C2/m$ CrI$_3$ phases and the Cr$_{2}$O$_{3}$ impurity phase, both for the nuclear (``str'') intensity alone and with the magnetic intensity (in the M-AFM model) included. The region of focus ($2.45 \leq d \leq 3.9$ \AA) is shaded. (b,c) Elastic intensity within (c) 70 to 200 K and (d) 200 to 275 K, with a linear background subtracted to account for its temperature dependence.}
\label{fig:2}
\end{figure}

Neutron scattering data on a sample of CrI$_3$ powder (which had been ground for a few minutes in a mortar and pestle) were taken on the VISION instrument at Oak Ridge National Laboratory, which collects elastic and inelastic data on two separate sets of detectors. 
The elastic neutron scattering data collected at 5 K is shown in Fig.\ \ref{fig:2}(a) as a function of $d$-spacing, where $d=\frac{2 \pi}{Q}$ and $Q$ is the momentum transfer. Also shown are the simulated intensities for the $R\bar{3}$ and $C2/m$ phases and a Cr$_{2}$O$_{3}$ impurity phase. As shown in Fig.\ \ref{fig:2}(b), there is minimal change in the intensity on warming from 70 to 200 K. Only localized changes are seen from 200 to 275 K (Fig.\ \ref{fig:2}(c)), in the form of the expected shrinking of $R\bar{3}$ peaks (such as $(113)_R$) and growing intensity of $C2/m$ peaks (such as $(131)_M$). 
Overall, though, it is clear that the intensity in the $2.7 \leq d \leq 3.5$ \AA\ range (highlighted in green in Fig.\ \ref{fig:2}(a)) cannot be represented by the ordered $R\bar{3}$ and $C2/m$ phases alone, and that substantial diffuse scattering is present arising from disordered R- and M-type layer stacking. 

The percentages of M- and R-type stacking was estimated from Rietveld refinement at low $d$/high $Q$ where the intensity of a randomly stacked R/M mixture can be approximated by a linear combination of intensity arising from the two phases ($R\bar{3}$ and $C2/m$; see Supplemental Section B). At 5 K, the sample consists of about 63\% M-type and 37\% R-type stacking. A Cr$_2$O$_3$ second phase is present as well at about 5 wt\%. There are additional, likely magnetic, peaks near $d=5.0$ \AA\ that arise below $\sim$20 K. We have not identified the source of these peaks, but note that low-energy ($\hbar \omega < 4$ meV) spin-wave intensity also appears below $\sim$20 K.

\begin{figure}[h]
\begin{center}
\includegraphics[width=8.6cm]
{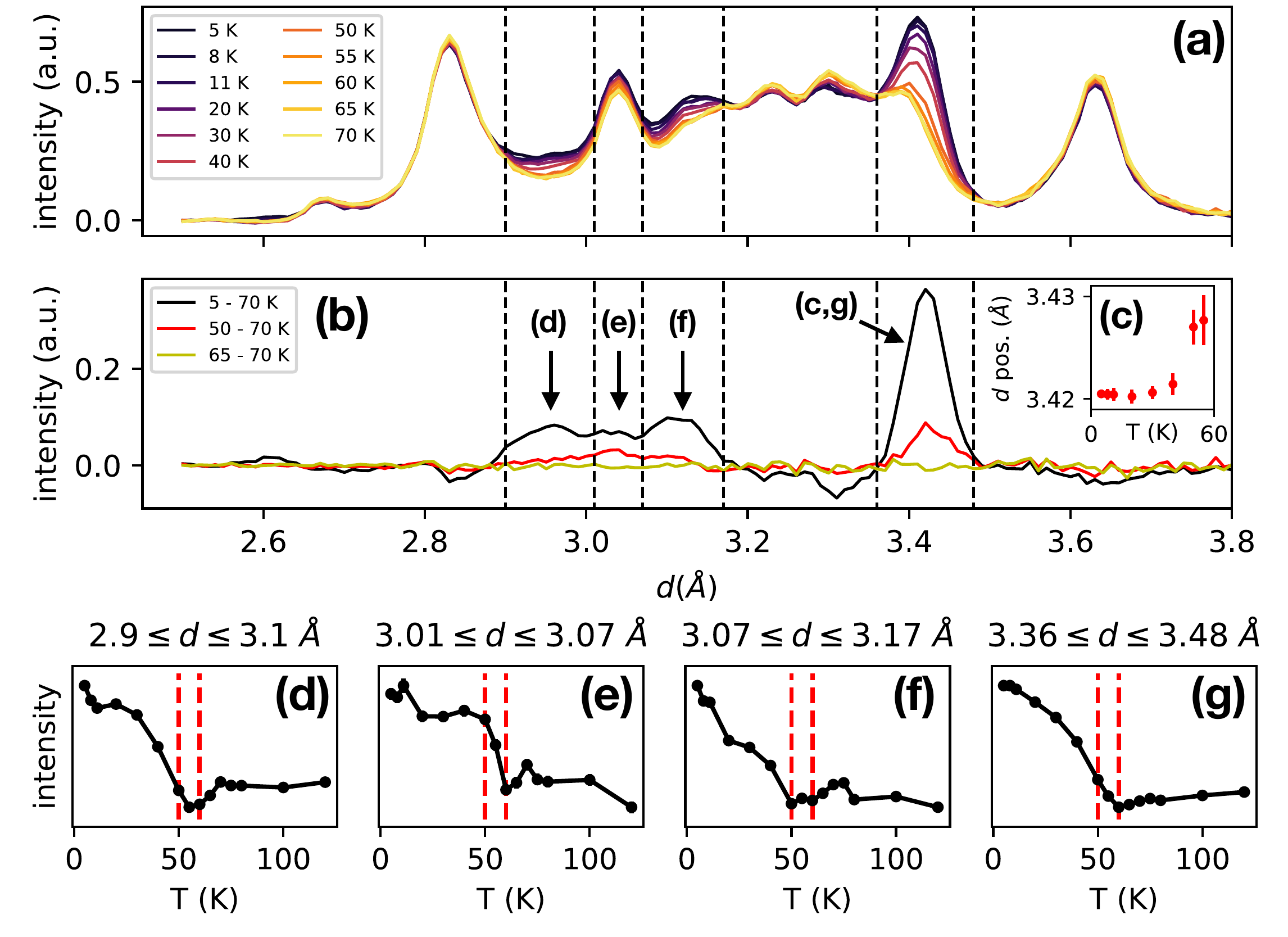}
\end{center}
\caption{(a) Elastic scattering intensity vs.\ layer spacing $d$. A linear background was subtracted for each temperature. (b) Intensity with the 70 K data subtracted for $T=5$, 50, and 65 K, to show the magnetic contribution. (c) Fitted position of the peak in (b) near $d=3.43$ \AA\ vs.\ temperature.  (d-g) Integrated intensity of the raw data within the regions labeled in (b), plotted vs.\ temperature; red dashed lines show 50 and 60 K.}
\label{fig:3}
\end{figure}

\begin{figure}[h]
\begin{center}
\includegraphics[width=8.6cm]
{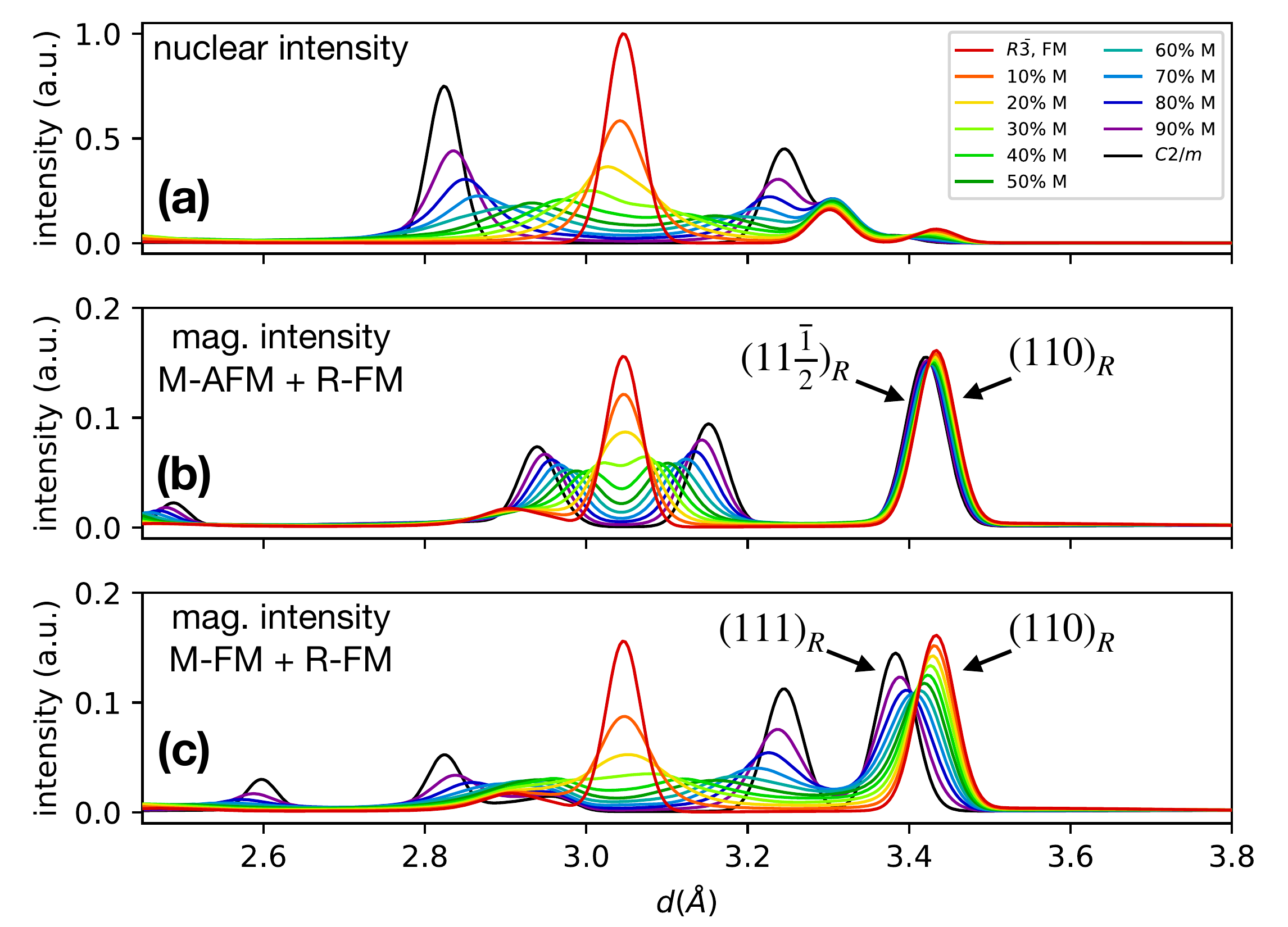}
\end{center}
\caption{Simulation of the (a) nuclear diffuse scattering intensity, and magnetic diffuse scattering intensity within the (b) M-AFM and (c) M-FM models for various percentages of M-type stacking.}
\label{fig:4}
\end{figure}

In Fig. \ref{fig:3}(a) the temperature dependence of the elastic intensity is shown from 5 to 70 K in the range of $2.5 \leq d \leq 3.8$ \AA. Since the intensity does not change from 70 to 200 K, we use the 70 K data as a background to subtract from the 5, 50, and 60 K data, leaving behind the magnetic intensity in Fig. \ref{fig:3}(b). A strong peak at $d=3.43$ \AA\ and broader intensity around $d=3.04$ \AA\ are present at 5 K. By 50 K, this intensity is diminished and changes shape around $d=3.04$ \AA, becoming more concentrated toward the center. 

To identify the origin of the structural diffuse scattering, the intensity from an R/M random stacking model was simulated, and the results are shown in Fig.\ \ref{fig:4}(a-c). The simulated intensity was obtained from the squared structure factor of a supercell constructed with a random mixture of R- or M-type stacking. From the simulated nuclear structural intensity in Fig.\ \ref{fig:4}(a), it is evident that R/M stacking disorder does, indeed, result in a broadening of the intensity within the $2.8 \leq d \leq 3.4$ \AA\ range. (We show in the Supplemental Materials (Section D) that, although there are two R-type and three M-type stacking options, the specific types of M- or R-type stacking involved have only a subtle effect on the intensity.) 
In Fig.\ \ref{fig:4}(b,c), we present two models for the magnetic scattering. The M-AFM model has flipped spins across every M-type stacking boundary (as depicted in Fig.\ \ref{fig:1}(b)), and the M-FM model assumes all of the spins are aligned in the same direction regardless of stacking. A cursory comparison between the results of these two models and the magnetic intensity in Fig.\ \ref{fig:3}(b) shows that the M-AFM model has much better agreement with our data. 

Strikingly, the M-AFM model predicts that the $(110)_R$ peak near $d=3.43$ \AA\ remains almost unchanged as R-type stacking is replaced with M-type stacking, with its $d$-spacing shifting by only -0.013 \AA\ from $(110)_R$ to the corresponding $C2/m$ peak at $(11\bar{\frac{1}{2}})_R$. (See Supplemental Section D for a mathematical explanation.) In Fig.\ \ref{fig:3}(c), we show the fitted position of this peak as a function of temperature, showing an abrupt shift above 50 K of about +0.007 \AA. If we assume this change corresponds to a shift toward $(110)_R$ from $(11L)_R$ where $L$ represents the average position of peaks arising from a distribution of M-type stacking fractions, we obtain an estimate of 73(8)\% M-type stacking, roughly consistent with our estimate of $\sim$63\% from the low-$d$ refinement. (A slight increase in the width of the $(110)_R$ peak below $\sim$53 K was reported in Ref.\ \cite{chenMagneticAnisotropyFerromagnetic2020} and interpreted as evidence that the spin-spin correlation length was finite even at low temperature, but in light of our results, such a peak broadening is, instead, likely due to the presence of a distribution of M-type stacking fractions in the sample, resulting in a superposition of peaks at $(11L)_R$ with a range of $L$ values within $-\frac{1}{2} \leq L \leq 0$.) 

If M-type stacking is associated with a transition at $\sim$50 K, then we would expect intensity associated with M-type stacking to decrease on warming faster than for R-type stacking. This is exactly what is seen in Figs.\ \ref{fig:3}(d-f), which are plots of the temperature dependence of the intensity integrated within the $d$ ranges indicated by the dashed lines in Fig.\ \ref{fig:3}(b). 
From the simulated M-AFM magnetic intensity (Fig.\ \ref{fig:4}(b)), it is clear that the intensity near $d=3.04$ \AA\ is disproportionately from R-type stacking, while the intensity near $d=2.96$ \AA\ and $3.12$ \AA\ is predominately from M-type stacking. The intensity at $d=2.96$ \AA\ and $d=3.12$ \AA\ show transitions at or just above 50 K, while the intensity at $d=3.04$ \AA\ shows a transition at $\sim$60 K. This change can also be seen in the (50-70) K data in Fig.\ \ref{fig:3}(b), in which a peak near $d=3.04$ \AA\ is still present but its two side peaks at $2.96$ and $3.12$ \AA\ are absent. Meanwhile, the peak near $d=3.43$ \AA\ has contributions from both M- and R-type stacking, and its intensity thus shows an ultimate transition at the higher of the two transition temperatures, $\sim$60 K (Fig.\ \ref{fig:3}(g)). Thus, our elastic neutron scattering data provide evidence for the magnetic coupling across M-type stacking boundaries arising below $\sim$50 to 55 K.

\begin{figure}[h]
\begin{center}
\includegraphics[width=8.6cm]
{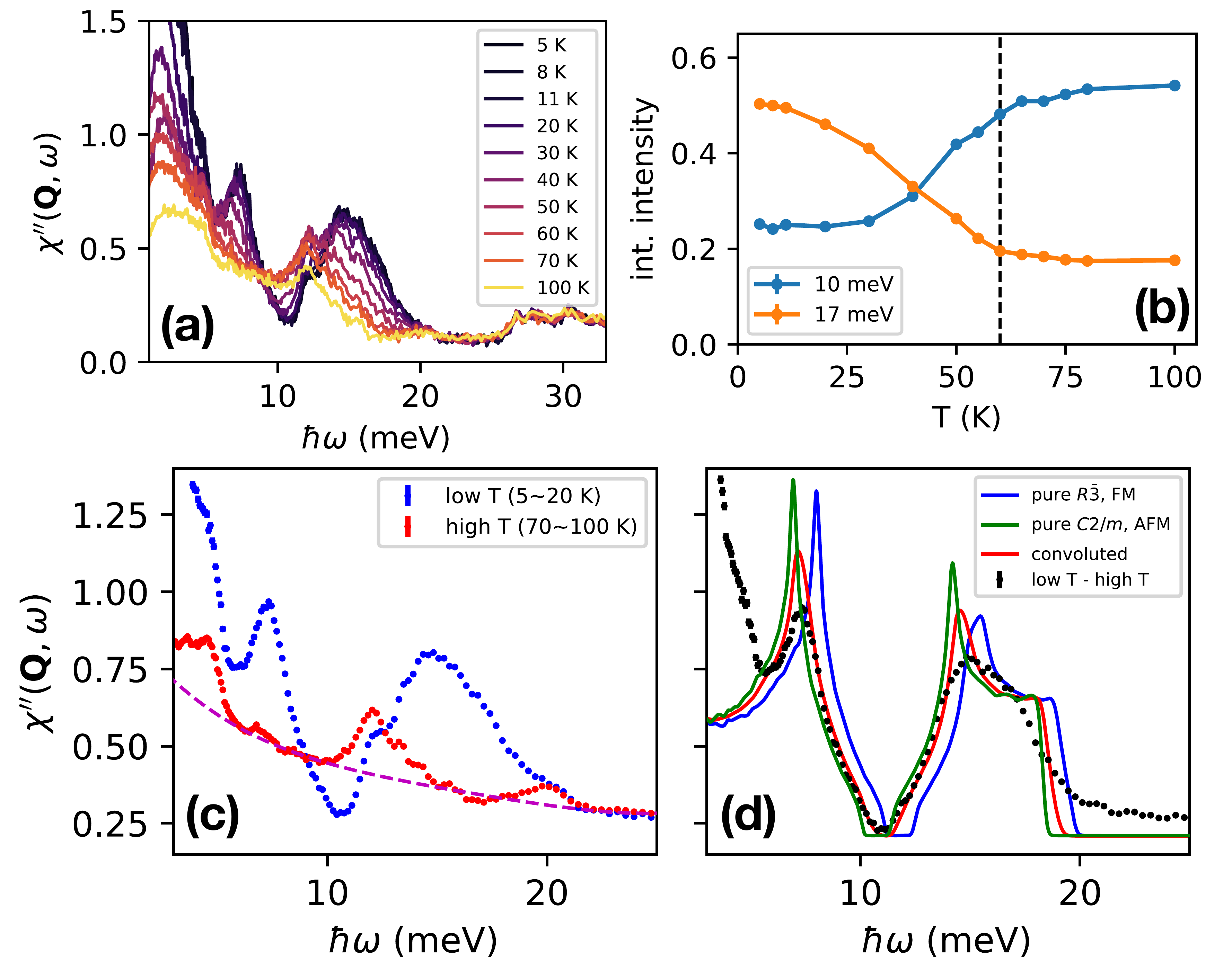}
\end{center}
\caption{(a) Inelastic neutron scattering intensity (Bose-factor corrected) as a function of energy transfer for temperatures taken on warming from 5 to 275 K along the low-$Q$ trajectory of VISION. (b) Temperature dependence of inelastic intensity near 10 and 17 meV, averaged within $\pm$0.5 meV. The dashed line indicates 60 K. (c) Inelastic intensity (Bose-factor corrected) plotted vs.\ energy transfer with averaging over two sets of data: ``low T'' (5, 8, 11, and 20 K), and ``high T'' (70, 75, 80, and 100 K). The dashed line is a polynomial background fit to the high-T data. (d) To account for phonon peaks, the high T data (with the fitted background subtracted) was subtracted from the low T data, and plotted as black points. Also shown are curves of simulated intensity for the ``$R\bar{3}$, FM'' model (calculated from the ``J-DM'' parameters in Ref.\ \cite{chenMagneticFieldEffect2021}), for the ``$C2/m$ AFM'' model (with a summed interlayer interaction of 0.073 meV \cite{cenkerDirectObservationTwodimensional2021,chenMagneticFieldEffect2021}, and shifted by -0.2 meV), and for a ``blended'' model where the $C2/m$ model was convoluted with a distribution of energy shifts assuming 67\% M-type stacking.}
\label{fig:5}
\end{figure}

\section*{Inelastic neutron scattering}
Inelastic neutron scattering intensity is shown in Fig. \ref{fig:5}(a). VISION collects inelastic data at fixed incident neutron energy along two sets of detector banks; we focus on the ``low-$Q$'' data set where the magnetic intensity is stronger. The spin-wave dispersion of CrI$_{3}$ resembles that of the electronic band structure of graphene, where acoustic and optic branches disperse along the in-plane directions and meet at Dirac points. 
The data indeed show acoustic- and optic-branch features at temperatures below $\sim$60 K, similar to CrCl$_3$ data also taken on VISION \cite{schneelochGaplessDiracMagnons2022}. 
The optic branch hump is centered around 15 meV, separated from the acoustic branch by a Dirac gap around 10 to 11 meV. A peak at the acoustic branch saddle-point can be seen at 7.3 meV, while there is a lack of a clear optic branch saddle-point peak, presumably due to broadening by interlayer interactions or mixed stacking. 
Below 4 meV, additional features are present at temperatures lower than 20 K, likely due to the magnetic impurity phase discussed above, but minimal change is observed above 4 meV in this temperature range. Spin waves would also arise from the Cr$_2$O$_3$ impurity phase, but the energy scale of the dispersion is higher, with maxima around 40 to 50 meV \cite{samuelsenInelasticNeutronScattering1970}, and the intensity is expected to be temperature-independent below $\sim$100 K since $T_N=308$ K for Cr$_2$O$_3$.

Little change is seen on warming (for $\hbar \omega \geq 4$ meV) until $\sim$30 K, at which point magnon dampening is observed, with the magnetic intensity being replaced by a paramagnetic background. These changes continue until $T_C \approx 60$ K, as seen from the temperature dependence of the intensity near 10 and 17 meV (integrated within $\pm$0.5 meV) in Fig.\ \ref{fig:5}(b). This temperature response is different from CrCl$_3$, where the spin-wave energy decreases continuously, even across the N\'{e}el transition \cite{schneelochGaplessDiracMagnons2022}; this different behavior is likely due to the interlayer magnetic coupling which is two orders of magnitude smaller in CrCl$_3$ \cite{narathSpinWaveAnalysisSublattice1965} than in CrI$_3$ \cite{chenMagneticFieldEffect2021}. At 10 meV, the intensity increases as the spin-wave renormalization fills in this energy range. At 17 meV, at the upper part of the optic branch, the intensity gradually decreases, and levels off at 60 K. Since the observed spin waves arise from the sample as a whole (i.e. from regions with R-type as well as M-type stacking), the presence of the transition at $\sim$60 K is as expected. 

Shown in Fig.\ \ref{fig:5}(c) is the dynamic susceptibility where the data from 5 to 20 K were averaged together to improve statistics (blue points). To remove the phonon contribution, 
data averaged from 70 - 100 K (red points), with a polynomial background fitted (magenta line) and subtracted, were subtracted from the 5 to 20 K data, as shown in Fig. \ref{fig:5}(d) (black points). Although the effect of stacking disorder will be considered below, there appears to be a gap around 11.0 meV that is roughly 1 meV wide.

The spin-wave intensity in Fig.\ \ref{fig:5}(d) is shifted downward by just under 1 meV relative to observed $R\bar{3}$-phase spin-wave energies, as represented by a calculation based on the ``J-DM'' model of Ref.\ \cite{chenMagneticFieldEffect2021}, which we plot as ``$R\bar{3}$, FM''. 
The calculated intensity in Fig.\ \ref{fig:5}(d) was obtained from a powder-averaged simulation in SpinW \cite{tothLinearSpinWave2015}, then convoluted with a narrow energy resolution \cite{seegerResolutionVISIONCrystalanalyzer2009} and a broad $Q$-resolution (assuming a FWHM spread in scattering angle of about 25$^{\circ}$, or $\sim$0.5 \AA$^{-1}$.) 
The ``$R\bar{3}$, FM'' model includes three in-plane exchange interactions, single-ion anisotropy, a Dzyaloshinskii-Moriya interaction, and two interlayer magnetic coupling constants for the 1st- and 2nd-nearest-neighbor interlayer Cr-Cr bonds. 
We use this model as representative of $R\bar{3}$-phase spin wave energies since it agrees well with data  \cite{chenMagneticFieldEffect2021,chenMasslessDiracMagnons2021}, at least for the locations of the saddle-point peaks and Dirac gap, though the model does disagree in the higher-energy region (seen in powder data \cite{chenMasslessDiracMagnons2021}) where it predicts a sharp dropoff in intensity while the data show a gradual decrease. 
Regardless, it is clear that there is a significant difference between the spin-wave energies in our data and those observed for $R\bar{3}$, likely due to the prevalence of M-type stacking in our sample. 
A similar energy shift ($\sim$0.5 meV) relative to $R\bar{3}$-phase expectations can be discerned in inelastic tunneling spectroscopy data on (presumably M-stacked) bilayer CrI$_3$ \cite{kimEvolutionInterlayerIntralayer2019}.
Interestingly, a $\sim$1 meV shift has also been observed in data on a powder CrI$_3$ sample that had been ball-milled overnight \cite{chenMasslessDiracMagnons2021}. However the elastic intensity for that sample was largely featureless, lacking the clear peaks of our data in Fig.\ \ref{fig:3}(a), suggesting that ball-milling overnight (rather than grinding for a few minutes in a mortar and pestle) led to a nearly amorphous crystal structure, well beyond the stacking disorder present in our sample. The inelastic features in our data are also much sharper than those observed for the ball-milled sample of Ref.\ \cite{chenMasslessDiracMagnons2021}, which were broadened well beyond resolution. 

The primary effect of changing the interlayer magnetic coupling is to apply an energy shift to the spin wave intensity, since the interlayer coupling is a small perturbation compared to intralayer interactions. We introduce the ``$C2/m$, AFM'' model, which has the same (intralayer) parameters as for ``$R\bar{3}$, FM'', except that the interlayer magnetic coupling, which sums to -0.59 meV per Cr$^{3+}$ ion for the $R\bar{3}$ model, is replaced with an AFM interlayer exchange of +0.073 meV (i.e., +0.073/4 meV per nearest-neighbor interlayer bond in the $C2/m$ structure.) The value of +0.073 meV is based on an analysis \cite{chenMagneticFieldEffect2021} of Raman spectroscopy data on bilayer CrI$_3$ \cite{cenkerDirectObservationTwodimensional2021}. 
In Supplemental Section H, we compare the $C2/m$ and $R\bar{3}$ models directly, and see that the main difference is an energy shift of $\sim$0.8 meV from $C2/m$ to $R\bar{3}$. The Dirac gap, in particular, remains largely unchanged. (We note that it is the \emph{magnitude} of the interlayer coupling which determines the size of the energy shift, since there is cancellation in simultaneously swapping the sign of the interlayer coupling constants and the directions of the spins.) 

For the mixed M/R stacking of our sample (with an estimated average of $\sim$63\% M-type stacking according to our low-$d$ refinement), we can approximate the expected inelastic intensity curve by convoluting the $C2/m$ curve with a function of the distribution of energy shifts, assumed to be linear in the distribution of R-type stacking fraction in the sample. We assume a stacking-fraction distribution that varies linearly from 100\% M-type to 0\% R-type stacking, having an average of 67\% M-type stacking. (Before performing the convolution, we shifted the $C2/m$ curve by -0.2 meV (as seen in Fig.\ \ref{fig:5}(d)) so that the resulting convoluted curve would line up with our data.) The result is plotted as the red curve in Fig.\ \ref{fig:5}(d). 

In both the data and the convoluted model intensity, a gap can be seen at the Dirac point. The effect of the convolution is to narrow the gap somewhat, but the gap's existence in our data appears to be a natural consequence of the gap being the same size in the ``$C2/m$, AFM'' and ``$R\bar{3}$, FM'' models. Thus, topological magnons may be present in the $C2/m$ phase, as has been proposed for the $R\bar{3}$. Such a result would not be surprising, since the gap is (presumably \cite{chenTopologicalSpinExcitations2018,chenMagneticFieldEffect2021}) opened by intralayer Dzyaloshinskii-Moriya interactions, which would remain about the same regardless of overall layer stacking.

\begin{figure}[h]
\begin{center}
\includegraphics[width=8.6cm]
{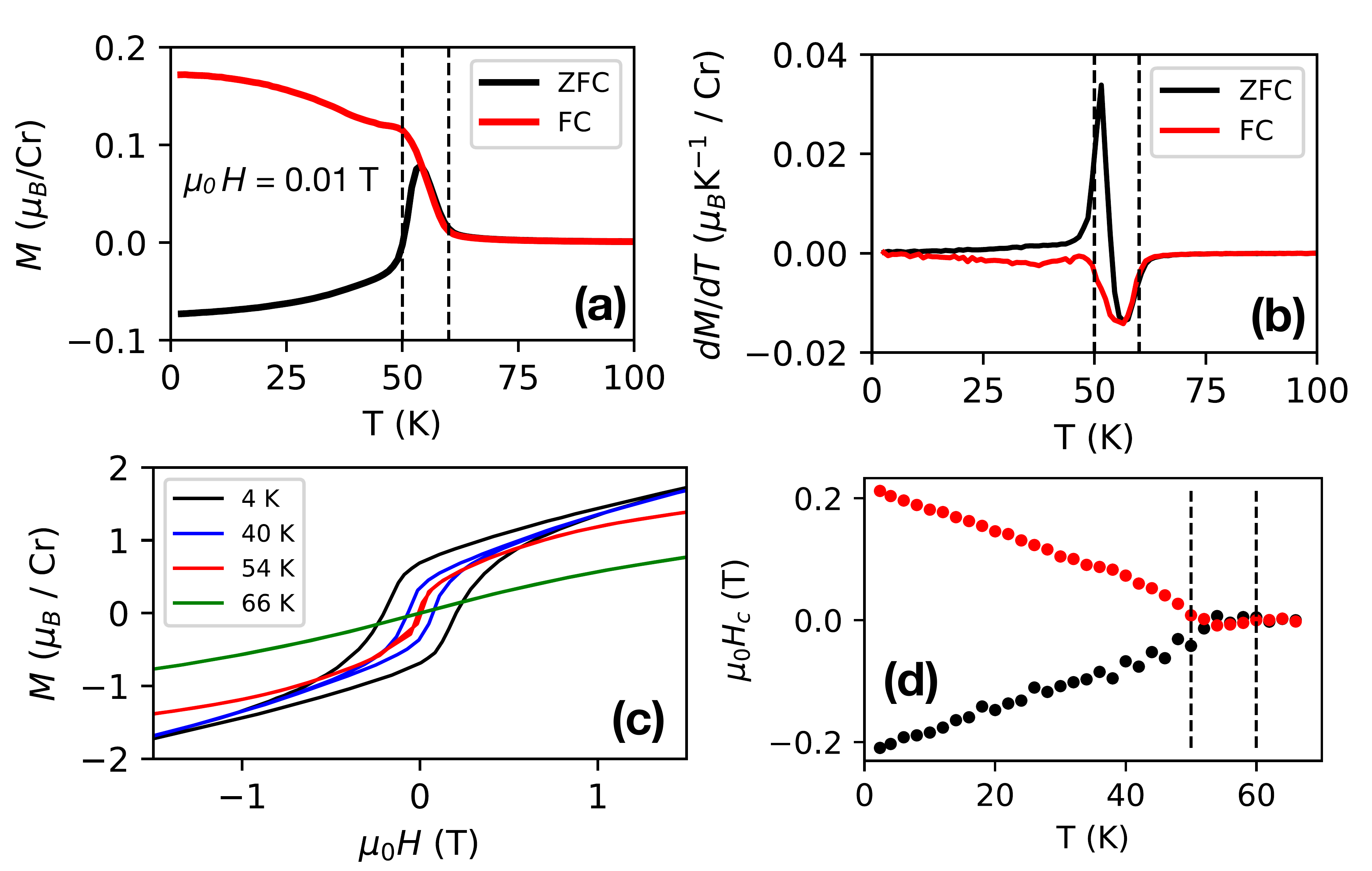}
\end{center}
\caption{(a) Magnetization vs.\ temperature for a pressed pellet of CrI$_3$ on warming (ZFC) and cooling (FC), taken at $\mu_0 H = 0.01$ T. (b) Derivative $dM/dT$ of the data in (a). (c) Magnetization-field hysteresis loops collected at several temperatures on a pellet of pressed CrI$_3$ powder. A hysteresis is present at 4 and 40 K, but is gone by 54 K. (d) The coercive field $\mu_0 H_c$ (the field at which $M=0$) plotted as a function of temperature, extracted from magnetization-field hysteresis loops. The hysteresis disappears around 52 K.}
\label{fig:6}
\end{figure}

\section*{Magnetization measurements on a pressed pellet}
Magnetization data also indicates a connection between stacking disorder and magnetic ordering. We performed magnetization measurements on a pressed pellet of ground CrI$_{3}$ powder, presumed to preserve disordered M-type stacking down to low temperature. In Fig.\ \ref{fig:6}(a), the magnetization $M$ as a function of temperature $T$ is shown, with its slope $dM/dT$ plotted in Fig.\ \ref{fig:6}(b). The sample was first cooled to 2 K, at which point a field of $\mu_0 H = +0.01$ T was applied, and zero-field-cooled (ZFC) data were collected on warming to 300 K. Field-cooled (FC) data were then collected on cooling back to 2 K. On warming, we see that the +0.01 T field is initially insufficient to reverse the sample's negative magnetization that happened to have set in on its first cooling. Above $\sim$50 K, however, the magnetization rises sharply and becomes positive. The AFM coupling across the M-type boundaries causes the spin direction direction to flip back and forth on crossing these boundaries, resulting in an almost random spontaneous magnetization in any given region, but above $\sim$50 K, the disappearance of the AFM coupling leaves disconnected FM R-type-stacked regions that are free to align in response to a small field. With higher temperature comes greater thermal fluctuations, and thus the magnetization in Fig.\ \ref{fig:6}(a) drops on further warming, with FM order vanishing near the usual transition temperature of $T_C=61$ K \cite{mcguireCouplingCrystalStructure2015}. 
On cooling, the magnetization rises sharply below $\sim$60 K, but flattens just under 50 K before having an upturn on further cooling, showing the resistance to full FM alignment induced by the AFM coupling of the M-type stacking.
The magnetization reaches a level of $\sim$0.17 $\mu_B$/Cr, comparable to values reported in the literature for single crystals with $\mu_0 H=0.01$ T applied out-of-plane \cite{liuAnisotropicMagnetocaloricEffect2018,mcguireCouplingCrystalStructure2015}, though those studies report the magnetization approaching its maximum near 60 rather than 50 K. 
(At much larger fields of $\pm$9 T, we observe full magnetic saturation near $\pm3 \mu_B$/Cr$^{3+}$ ion; see Supplemental Section I.)

Perhaps the clearest signal of the AFM transition in the magnetization data is the disappearance of a magnetization-field hysteresis loop above $\sim$50 K. These data are shown in Fig.\ \ref{fig:6}(c) for selected temperatures, and the coercive field $\mu_0 H_c$ (i.e., $\mu_0 H$ where $M=0$) is plotted in Fig.\ \ref{fig:6}(d). Despite the presence of FM order up to $\sim$60 K, the hysteresis vanishes around 52 K, showing the role of the AFM coupling across M-type stacking boundaries in pinning the magnetization. 
(Metamagnetic jumps on increasing out-of-plane field past 2 T have been reported \cite{liuThicknessdependentMagneticOrder2019}, attributed to the collapse of interlayer AFM across M-type stacking boundaries, but we have not observed these features, likely due to the polycrystalline nature of our sample.) 
Anomalies in magnetization vs.\ temperature data on bulk crystals have been reported before \cite{meseguer-sanchezCoexistenceStructuralMagnetic2021,arnethUniaxialPressureEffects2022,liuThicknessdependentMagneticOrder2019}, sometimes attributed to AFM ordering across M-type stacking but usually with the assumption that the behavior is confined to surface layers \cite{niuCoexistenceMagneticOrders2020}. An increase in the coercive field with decreasing crystal thickness was also reported \cite{liuThicknessdependentMagneticOrder2019}, though not directly attributed to M-type stacking. 
However, to our knowledge, the disappearance of the magnetization-field hysteresis above $\sim$50 K has not been reported before, even as it makes clear the connection between the magnetic anomalies and the presence of AFM order across M-type stacking boundaries. The hysteresis may have practical applications (e.g., since mixed stacking evidently induces hysteresis, it may be a strategy for improving retentivity in data storage based on vdW-layered magnetic materials \cite{zhangHardFerromagneticBehavior2022}), but the hysteresis also provides a convenient way of diagnosing possible mixed magnetic ordering in other vdW-layered compounds where the type of magnetic order is correlated with stacking, such as Fe$_{5-x-y}$Co$_y$GeTe$_2$ \cite{mayTuningMagneticOrder2020}.

\section*{Discussion}
We have shown that, at low temperature, there is AFM order in CrI$_3$ wherever M-type stacking happens to be present. Despite the impression that the literature might bring, the link between AFM and M-type stacking is not limited to thin flakes 
or the surfaces of bulk crystals, and is very likely the source of anomalies in magnetization data \cite{liuThicknessdependentMagneticOrder2019} and the secondary phase seen via muon spin rotation \cite{meseguer-sanchezCoexistenceStructuralMagnetic2021}. 
The hypothesis that M-type stacking tends to be present at the surface of otherwise ideal $R\bar{3}$-structure crystals may be true \cite{niuCoexistenceMagneticOrders2020}, but even bulk crystals often have inhibited $C2/m$$\rightarrow$$R\bar{3}$ transitions, and the possibility of M-type stacking affecting magnetization behavior should not be overlooked. 

More generally, our results show that neutron scattering can uncover details about interlayer magnetism at the nanoscale, without the effort of preparing and measuring samples at the few-layer limit. Furthermore, as we can see from the closing of the isothermal magnetization loop above $\sim$50 K in CrI$_3$, magnetization measurements are a convenient way of obtaining essential hints as to the nature of interlayer magnetism. 

Beyond CrI$_3$, the effects of mixed interlayer magnetic coupling may be seen in many other compounds.
In CrCl$_3$, for instance, M-type stacking reportedly has a tenfold-greater interlayer AFM magnetic coupling than the usual R-type stacking \cite{kleinEnhancementInterlayerExchange2019}, 
but the potential of mixed stacking as a source of certain magnetization anomalies seen at low magnetic field \cite{bykovetzCriticalRegionPhase2019} has not been investigated. 
In CrBr$_3$, while the $M$$\rightarrow$$R$ structural transition is well above room temperature \cite{morosinRayDiffractionNuclear1964} and even few-layer flakes tend to be R-stacked \cite{hanAtomicallyUnveilingAtlas2023}, a kink in magnetization data \cite{yanSelfselectingVaporGrowth2023} suggests the possibility of AFM order across M-type boundaries in CrBr$_3$. 
Cr$_2$Si$_2$Te$_6$ and Cr$_2$Ge$_2$Te$_6$ have also been reported to have anomalies in their magnetization data, attributed to magnetic anisotropy \cite{xieTwoStageMagnetization2019}, but the possibility of mixed stacking should not be discounted. RuCl$_{3}$ is another honeycomb-layered material that is structurally similar to the chromium trihalides; it also has multiple magnetic transitions associated with stacking defects \cite{loidlProximateKitaevQuantumspin2021} (e.g., deforming a crystal introduces a second magnetic transition \cite{caoLowtemperatureCrystalMagnetic2016}.) Finally, Fe$_{5-x}$GeTe$_2$, with $T_C \approx 310$ K, also reportedly has changes in both magnetic order and layer stacking as a function of Co doping \cite{mayTuningMagneticOrder2020}. 
In these materials, an analysis of diffuse neutron scattering intensity (and a careful look at magnetization data) in stacking-disordered samples may clarify the interlayer magnetic coupling. 
While the versatility of the interlayer magnetic coupling in CrI$_3$ was first discovered in nanoscale measurements, we can imagine that neutron scattering of ground powder samples may allow the discovery of this magnetic versatility in many other vdW-layered magnetic compounds before measurements with few-layer samples are attempted.

%\section{Conclusion}

In conclusion, we have performed elastic and inelastic neutron scattering measurements on a ground-powder CrI$_3$ sample. An analysis of the nuclear and magnetic diffuse scattering allows us to conclude that AFM spin alignment occurs across M-type stacking defects at temperatures below $\sim$50 to 55 K, even as FM order persists up to $\sim$60 K. Inelastic measurements showed a $\lesssim$1 meV decrease in spin-wave energy relative to a reported $R\bar{3}$-phase model, indicating that the magnitude of magnetic coupling across M-type boundaries is significantly less than across R-type boundaries. Magnetization measurements showed a hysteresis in the magnetization loop that vanishes above $\sim$52 K, implying that M-type stacking is responsible for pinning the FM domains.

\section*{Acknowledgements}

The authors would like to acknowledge discussions with Jeffrey Teo. The work at the University of Virginia is supported by the Department of Energy, Grant number DE-FG02-01ER45927. The Spallation Neutron Source is a DOE Office of Science User Facility operated by Oak Ridge National Laboratory. 

\section*{Methods}
Stoichiometric amounts of Cr and I powders were sealed into ampoules. The ampoules were heated at 100 $^{\circ}$C/h to 650 $^{\circ}$C, then kept there for three days before cooling to room temperature. The powder was ground for several minutes in a mortar and pestle (inside of an argon glove bag) before being put into a can and shipped to Oak Ridge National Laboratory for neutron scattering measurements. For magnetic susceptibility measurements, the powder was synthesized similarly, then pressed into a pellet while in an argon atmosphere. 

Neutron scattering measurements were taken on the VISION instrument at the Spallation Neutron Source at Oak Ridge National Laboratory. VISION is an indirect-geometry time-of-flight spectrometer. The final neutron energy was fixed at 3.5 meV. Inelastic data were taken on two detector banks at scattering angles of 45 and 135$^{\circ}$. Simultaneously, elastic data were taken on six detector banks at a 90$^{\circ}$ scattering angle. Unless otherwise indicated, the elastic data are the average of those of the six banks. The CrI$_3$ sample was cooled to 5 K, warmed to 140 K, cooled to 15 K, and warmed to 275 K; the data shown are from the warming portions (5 to 140 K, and 175 to 275 K.) Positions in reciprocal space that are labeled $(hkl)_R$ or $(hkl)_M$ correspond to $R\bar{3}$- or $C2/m$-phase reciprocal space coordinates, respectively. 

Magnetization measurements were performed in a Quantum Design Physical Property Measurement System equipped with a Vibrating Sample Magnetometer. 

\bibliography{CrI3Paper}

\section*{Supplemental Materials}
\beginsupplement

\subsection{Single-crystal x-ray diffraction showing inhibited transition}

\begin{figure}[h]
\begin{center}
\includegraphics[width=8.6cm]
{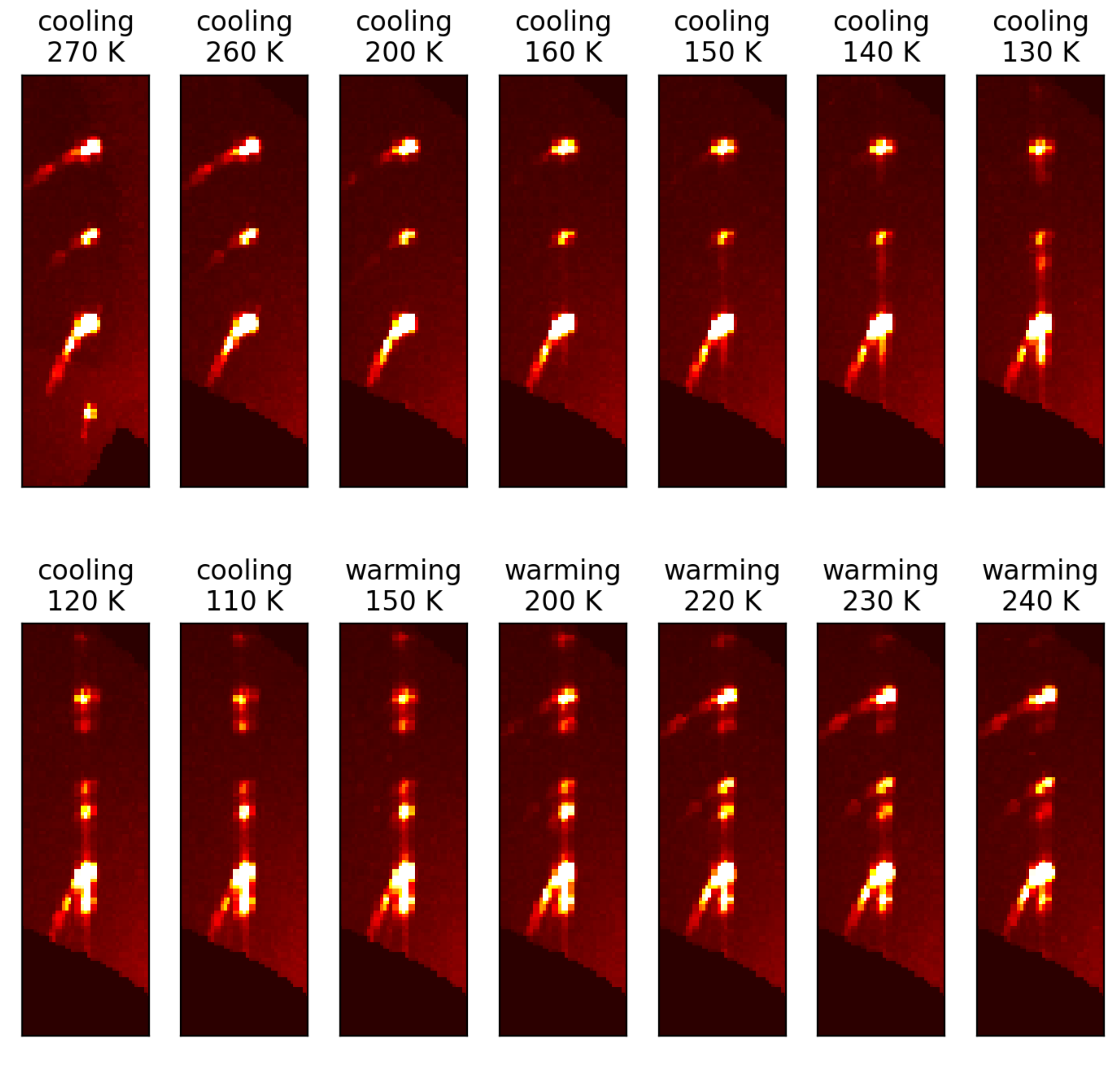}
\end{center}
\caption{Single-crystal X-ray diffraction images showing intensity along $(11L)_R$. Data taken on cooling from 270 to 110 K, then warming to 240 K.}
\label{fig:SingleCrystalXRD}
\end{figure}

Our single-crystal x-ray diffraction data shows an example of an inhibited $C2/m$$\rightarrow$$R\bar{3}$ transition. At high temperature, the $C2/m$ peaks in Fig.\ \ref{fig:SingleCrystalXRD} are located at $(11L)_R$ for $L=(1+3n)/3$ with integer $n$. (Along this line, the intensity would be the same for all three of the $C2/m$ twins, which are related by 120$^{\circ}$ rotations about the out-of-plane axis.) Around 160 K, a faint streak of diffuse scattering is seen, and $R\bar{3}$ peaks (at $L=3n$ for integer $n$) are seen soon after, but the transition is not complete by 110 K, the lowest accessible temperature during our measurement. On subsequent warming, the transition back to $C2/m$ begins around 160 K, but is not complete by 240 K. This behavior resembles that seen by McGuire, \emph{et al}.\ \cite{mcguireCouplingCrystalStructure2015}. Similar behavior has been seen in CrCl$_3$ \cite{mcguireMagneticBehaviorSpinlattice2017, schneelochGaplessDiracMagnons2022}.

\subsection{Estimated percentage of M- and R-type stacking}

\begin{figure}[h]
\begin{center}
\includegraphics[width=8.6cm]
{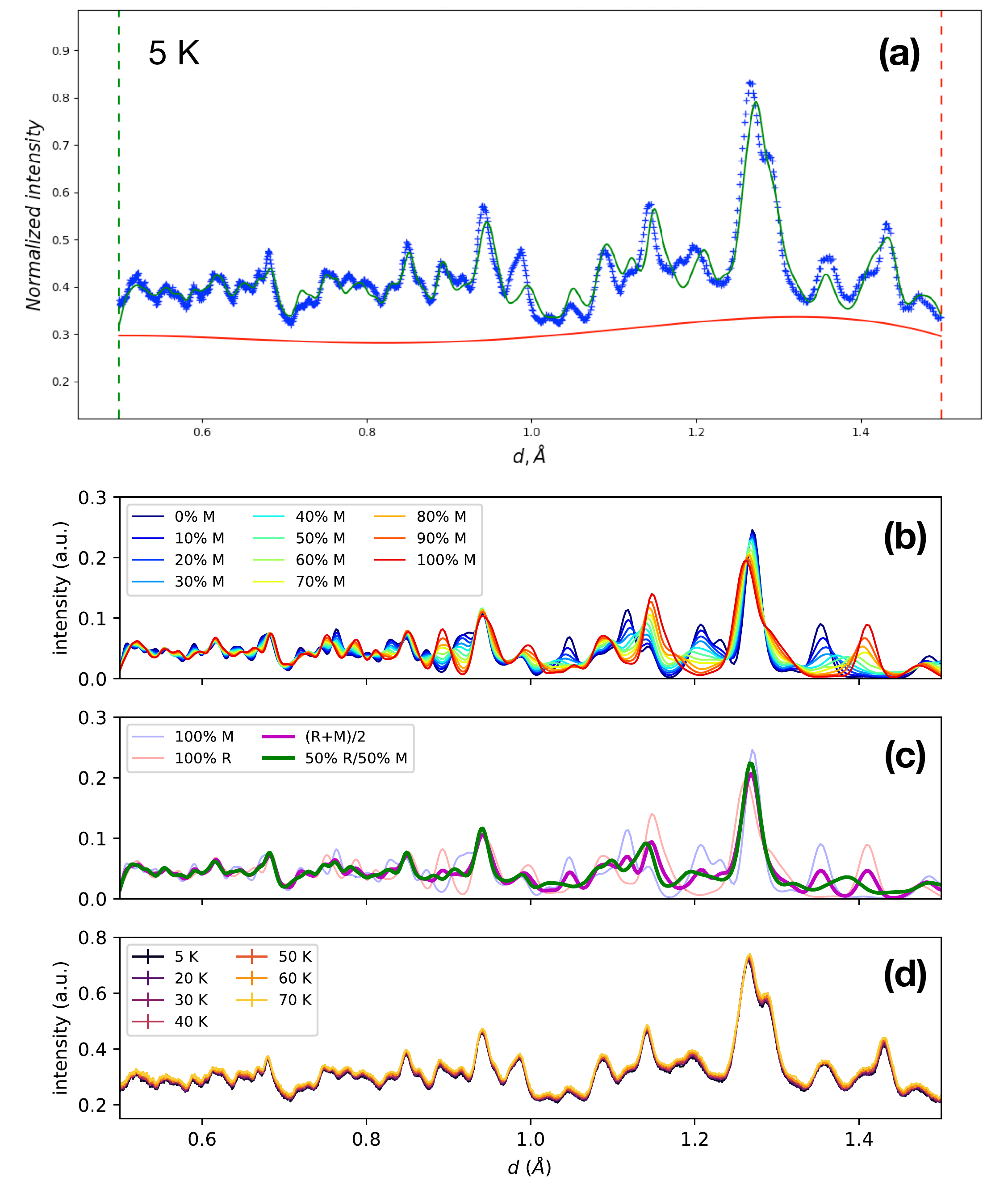}
\end{center}
\caption{(a) Result of Rietveld refinement of the 5 K elastic data (for one particular detector bank) within $0.5 \leq d \leq 1.5$ \AA\ in GSAS-II. Blue markers show the data, and the green curve shows the fitted intensity arising from 35.1 wt\% $R\bar{3}$ CrI$_{3}$, 60.0 wt\% $C2/m$ CrI$_{3}$, and 4.9 wt\% Cr$_{2}$O$_{3}$. For CrI$_3$ alone, these percentages are 36.9\% $R\bar{3}$ and 63.1\% $C2/m$. The red curve is the fitted background. (b) Simulated stacking intensity for a random stacking model for various percentages of M-type stacking vs.\ R-type stacking. (c) Simulated intensities, comparing 100\% M-type stacking, 100\% R-type stacking, the average of the two, and a 50\%/50\% random stacking mixture. (d) Elastic neutron scattering data on warming from 5 to 70 K, showing no signs of structural change.}
\label{fig:LowD}
\end{figure}

In this section, we estimate the percentage of M- and R-type stacking from a Rietveld refinement of data in the low-$d$/high-$Q$ region with respect to the $R\bar{3}$, $C2/m$, and Cr$_2$O$_3$ phases. 

In Fig.\ \ref{fig:LowD}(a), we compare the elastic intensity with results from refinement in GSAS-II \cite{tobyGSASIIGenesisModern2013}. The $R\bar{3}$- and $C2/m$-phase CrI$_3$ parameters were taken from the data reported in Ref.\ \cite{mcguireCouplingCrystalStructure2015} (for 90 and 250 K, respectively), except that, for $C2/m$, to compensate for thermal expansion, the $a$-axis lattice constant was set to that for $R\bar{3}$, $b$ was set to $\sqrt{3} a$, and $c$ was adjusted to match the interlayer spacing of $R\bar{3}$. (For simplicity, we neglect the $\sim$0.3\% difference in interlayer spacing between the two phases \cite{mcguireCouplingCrystalStructure2015}.) For Cr$_2$O$_3$, the parameters were taken from \cite{newnhamRefinementAlphaAl2O31962}. We accounted for the AFM ordering of Cr$_2$O$_3$ \cite{hillCrystallographicMagneticStudies2010} and set the magnetic moment on the Cr$^{3+}$ ions to 3 $\mu_B$. The refinement yielded 35.1(1) wt\% $R\bar{3}$ CrI$_{3}$, 60.0(1) wt\% $C2/m$ CrI$_{3}$, and 4.9(3) wt\% Cr$_{2}$O$_{3}$, implying that our sample had 63.1\% M-type stacking at 5 K. Refinement for the 70 K data resulted in similar values (35.7 wt\% $R\bar{3}$, 60.1 wt\% $C2/m$, and 4.3 wt\% Cr$_2$O$_3$), indicating that negligible structural change was seen on warming from 5 to 70 K (as one can also see from Fig.\ \ref{fig:LowD}(d).)

For a justification for our assumption that a refined 63\% volume fraction of $C2/m$ (vs.\ $R\bar{3}$ CrI$_3$) implies a roughly 63\% mixture of M-type (vs.\ R-type) stacking on average, we show simulations for a random stacking model in Figures \ref{fig:LowD}(b) and (c). The same R/M random stacking model was used as is discussed in the main text, and we will discuss extensions of this model (and how they have little effect on the intensity) in a later section. In Fig.\ \ref{fig:LowD}(b), the simulated intensity for a range of percentages of M-type stacking are shown. In Fig.\ \ref{fig:LowD}(c), we compare the intensity for 100\% M-type stacking, 100\% R-type stacking, the average of the two, and a 50\%/50\% random stacking mixture. We see that the 50/50 intensity is close to that of the average of R- and M-type stacking alone, in contrast to higher $d$ (i.e., in the $2.5 \leq d \leq 3.8$ \AA\ range of Fig.\ \ref{fig:3}.)

\subsection{Unknown magnetic impurity phase}
\begin{figure}[h]
\begin{center}
\includegraphics[width=8.6cm]
{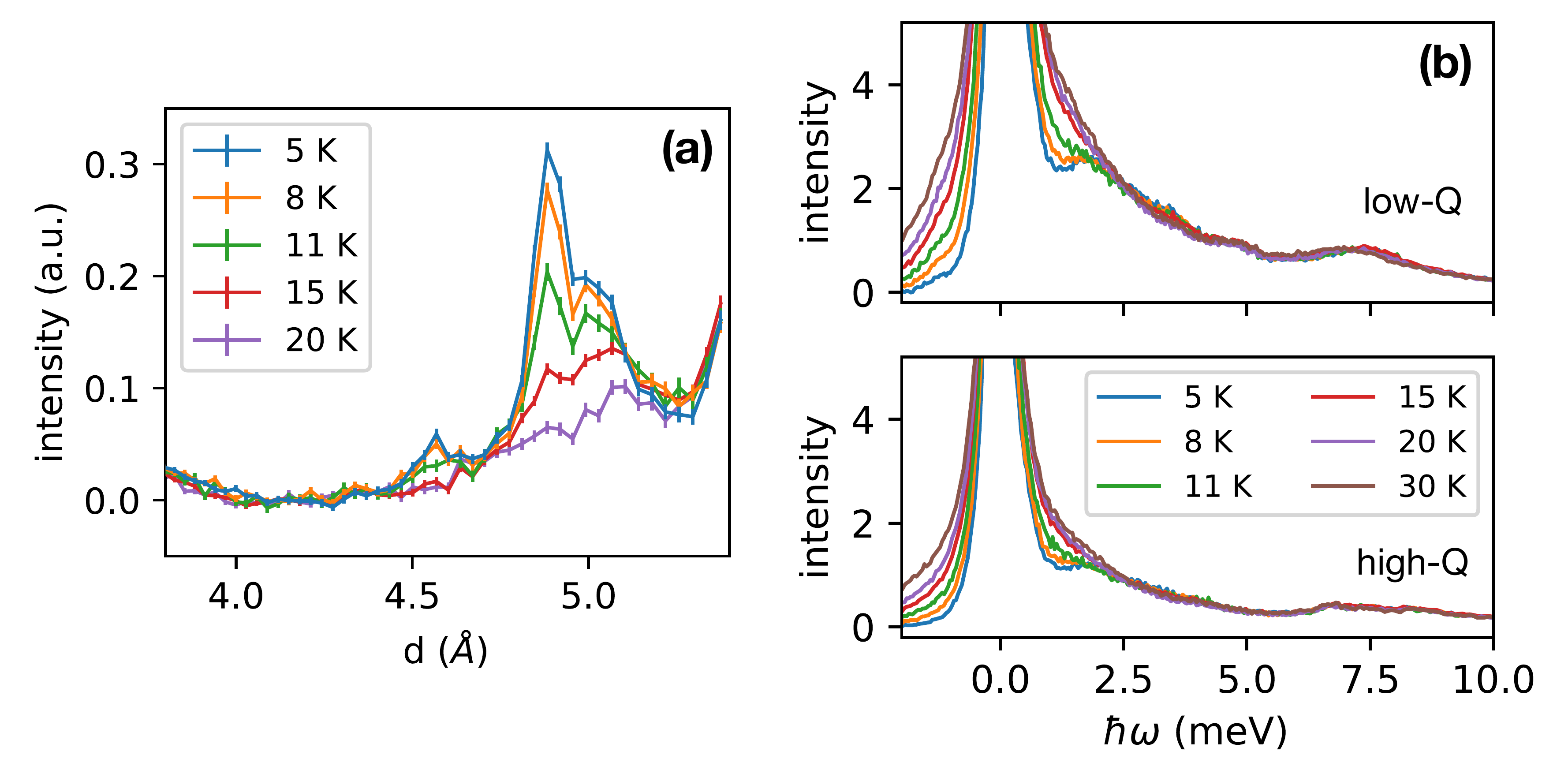}
\end{center}
\caption{(a) Elastic data showing the presence of high-$d$/low-$Q$ peaks near $d=4.54$ \AA, 4.89 \AA, and 5.0 \AA\ that vanish above 20 K. For each temperature, an offset of the average intensity within $4.1 \leq d \leq 4.2$ \AA\ was subtracted. (b) Inelastic data along the low-$Q$ and high-$Q$ trajectories of the VISION instrument showing a peak near 1.7 meV that vanishes above 20 K. Data were taken on warming except that at 15 K, which was collected on a later cooling run.}
\label{fig:impurityPhase}
\end{figure}

In addition to CrI$_3$ (with either R- or M-type stacking) and Cr$_2$O$_3$, there are signs of an additional magnetic ordering appearing below $\sim$20 K in our neutron scattering data, likely from an impurity phase. 
In Fig.\ \ref{fig:impurityPhase}(a), our elastic neutron scattering data show peaks near $d=4.54$ \AA, 4.89 \AA, and 5.0 \AA\ that vanish above 20 K; no other such peaks were found. The presence of these peaks exclusively at low-$Q$/high-$d$ and their appearance at low temperature suggests that they are magnetic in origin. In Fig.\ \ref{fig:impurityPhase}(b), inelastic data is shown, with a peak present near 1.7 meV. This intensity is larger along the low-$Q$ than the high-$Q$ trajectory, as expected for spin-wave intensity. This intensity vanishes above 20 K, transforming into a paramagnetic background. We have observed similar features in a separate VISION data set CrBr$_3$ below 20 K, but with different magnetic peak locations and spin wave energies. Since CrBr$_3$ and CrI$_3$ both absorb water from the air, and turn to liquid within days, we speculate that exposure to water vapor during synthesis or shortly afterwards may have produced an impurity phase containing hydrogen, chromium, either I or Br, and possibly oxygen in our CrI$_3$ and CrBr$_3$ samples, but a search through known compounds with these elements has not produced a match with our elastic neutrons scattering data. Alternatively, there could be an additional type of CrI$_3$ layer stacking, such as a type with alternating layer orientation as has been reported in bilayer CrBr$_3$ \cite{chenDirectObservationVan2019a}. 
We note that the muon spin rotation measurements of Ref.\ \cite{meseguer-sanchezCoexistenceStructuralMagnetic2021} report a third magnetic component (beyond the usual FM order and a second component associated with M-type stacking) below $\sim$25 K.

\subsection{Mathematical details of diffuse scattering simulation}

To compute the diffuse scattering, we calculated the squared structure factors (nuclear and magnetic) of a many-layer supercell with a random sequence of stacking options, and convoluted these Bragg peak intensities with a resolution. For our discussion, we use $R\bar{3}$-phase coordinates, with $\mathbf{a}_1 = (a,0,0)$, $\mathbf{a}_2 = (-\frac{a}{2},\frac{\sqrt{3} a}{2},0)$, and $\mathbf{a}_3 = (0,0,c)$, with the corresponding reciprocal lattice vectors labeled $\mathbf{b}_1$, $\mathbf{b}_2$, and $\mathbf{b}_3$. The two R-type stacking displacements (up to translational symmetry) can be represented as $\pm(\frac{1}{3} \mathbf{a}_1 + \frac{2}{3} \mathbf{a}_2) + \frac{1}{3} \mathbf{a}_3$, while the M-type stacking displacements are given by $-\frac{\alpha}{3} \mathbf{a}_1 + \frac{\epsilon}{3} \mathbf{a}_3$, $-\frac{\alpha}{3} \mathbf{a}_2 + \frac{\epsilon}{3} \mathbf{a}_3$, and $\frac{\alpha}{3} (\mathbf{a}_1 + \mathbf{a}_2) + \frac{\epsilon}{3} \mathbf{a}_3$, where $\alpha \approx 0.97$ and $\epsilon$ (the slight increase in interlayer spacing for M-type stacking vs.\ R-type stacking \cite{mcguireCouplingCrystalStructure2015}) is about $1.003$. The deviation of $\alpha$ or $\epsilon$ from 1 did not result in significant changes to our simulated intensity. For the intralayer atomic positions, we used those of the $C2/m$ phase. The reported differences in intralayer atomic positions \cite{mcguireCouplingCrystalStructure2015} between the three twins of $C2/m$ or the $R\bar{3}$ phase is on the order of $\sim$0.5\% of the in-plane lattice constants, and attempting to correct for the difference resulted in no significant change to simulated intensity. 

In the supercell, atomic positions ranged from 0 to 1 in the in-plane directions, but ranged from 0 to $N+1$ in the out-of-plane direction, where $N$ is the number of layers. Thus, the reciprocal lattice vectors $\mathbf{G} = H \mathbf{b}_1 + K \mathbf{b}_2 + L \mathbf{b}_3$ would have integer $H$ and $K$, but $L$ would be a multiple of $1/N$. In the limit $N \rightarrow \infty$, the diffuse scattering would vary continuously along $L$. 

For the (nuclear) structural intensity, the static nuclear structure factor is given by \cite{shiraneNeutronScatteringTripleaxis2002}
\begin{equation}
\label{eq:StrFac}
F (\mathbf{G}) = b_j e^{i \mathbf{G} \cdot \mathbf{d}_j}.
\end{equation}
The static magnetic structure factor is given by 
\begin{equation}
\label{eq:MagStrFac}
\mathbf{F}_{mag} (\mathbf{G}) = \sum_j p_j \mathbf{S}_{\perp j} e^{i \mathbf{G} \cdot \mathbf{d}_j}.
\end{equation}
For both equations, we assume the Debye-Waller factor is about unity in the region of interest and neglect it. $\mathbf{G}$ is a reciprocal lattice vector of the supercell. The index $j$ runs over every atom in the supercell for Eq.\ \ref{eq:StrFac}, and over the spins on the Cr atoms for Eq.\ \ref{eq:MagStrFac}. The nuclear scattering length is $b_j$. The vector $\mathbf{S}_{\perp j}$ is the component of the spin vector perpendicular to $\mathbf{G}$. The quantity $p_j = \frac{\gamma r_0}{2} g f(\mathbf{G})$, where $\gamma$ is the gyromagnetic ratio of the neutron, $r_0$ is the classical electron radius ($\frac{\gamma r_0}{2} = 2.695$ fm), $g$ is the Landé splitting factor, and $f(\mathbf{G})$ is the magnetic form factor. For our calculations, we assumed $g=2$ and $S=3/2$ for each Cr$^{3+}$ ion. 

\begin{figure}[h]
\begin{center}
\includegraphics[width=8.6cm]
{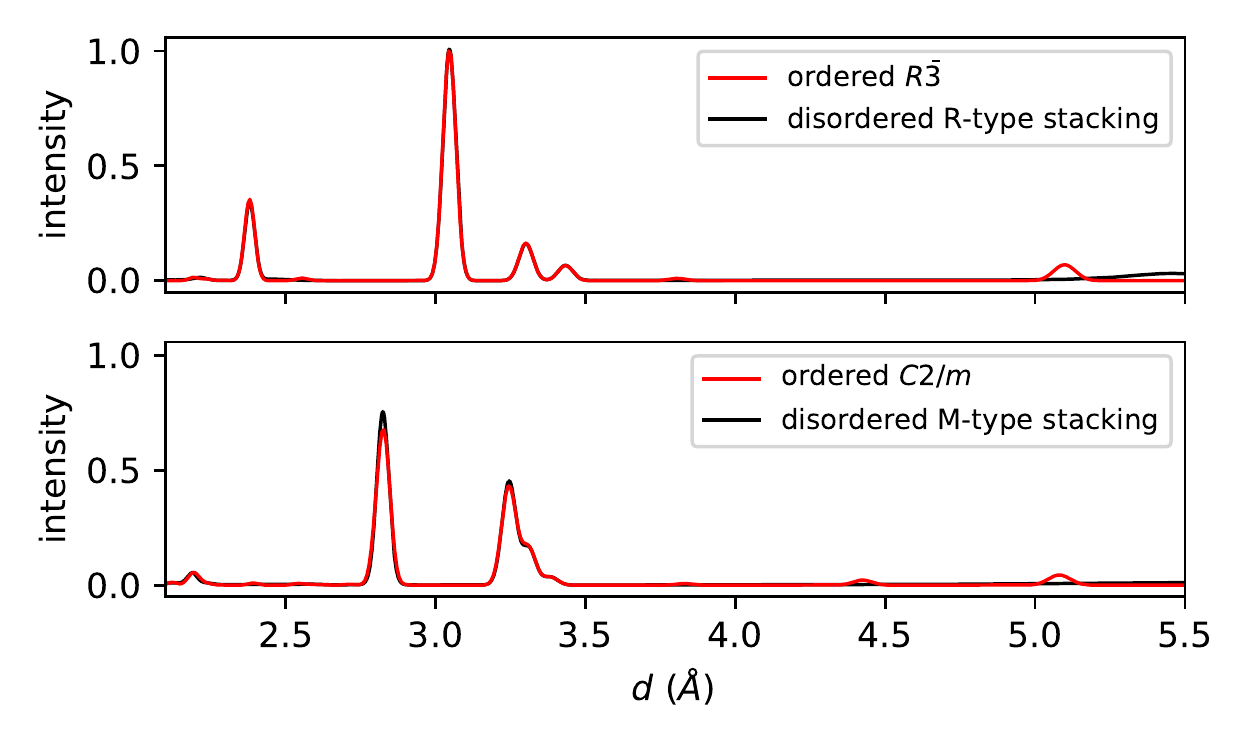}
\end{center}
\caption{Simulated intensity for perfectly-ordered stacking for $C2/m$ and $R\bar{3}$ phases, compared with a disordered sequence of the M- or R-type stacking, respectively.}
\label{fig:orderVsDisorder}
\end{figure}

For the diffuse scattering simulation in Fig.\ \ref{fig:4}, a supercell of 1920 layers was used. For Fig.\ \ref{fig:LowD}, a supercell of 240 layers was used. The simulated stacking sequence had random mixtures within the M- or R-type stacking; for example, a simulated curve labeled ``20\% M'' would consist of 6.7\% (20\%/3) of each of the three M-type stacking options, and 40\% (80\%/2) of the two R-type stacking options. Generally, stacking disorder that is exclusively M-type or R-type results in very weak diffuse scattering, as seen in Fig.\ \ref{fig:orderVsDisorder}. 

\subsection{Discussion of elastic peak near $d=3.4$ \AA.}

A key part of our argument that the spin direction flips wherever there is monoclinic-type stacking relies on the observation that our data shows a magnetic elastic peak at $d\approx3.4$ \AA\ rather than a broader feature. Here, we detail mathematically the source of this peak and why it manifests differently in the M-AFM model than the M-FM model.

The magnetic structure factor for the supercell can be rewritten as
\begin{equation}
\label{eq:LayerStrFac}
\mathbf{F}_{mag} (\mathbf{G}) = \sum_l s_l \mathbf{F}_l (\mathbf{G}) e^{i \mathbf{G} \cdot \mathbf{\Delta}_l}
\end{equation}
where $l=0...N$ is an integer indexing each layer, 
$s_l = \pm 1$ parameterizes the direction of the spins on layer $l$, 
$\mathbf{\Delta}_l$ is the overall translation vector for layer $l$,
and $\mathbf{F}_{mag,l} (\mathbf{G})$ is the magnetic structure factor for a single layer with $\mathbf{\Delta}_0 = (0,0,0)_R$ and spins aligned along $+\mathbf{\hat{a}_3}$. 

First, we consider a supercell with a random mixture of R-type stacking, so that $\mathbf{\Delta_l} = \frac{n_1 - n_2}{3} \mathbf{a}_1 + \frac{2 (n_1 - n_2)}{3} \mathbf{a}_2 + \frac{n_1 + n_2}{3} \mathbf{a}_3$, where $n_1$ and $n_2$ denote the number of each kind of R-type stacking option. Near $d=3.4$ \AA, there are six Bragg peaks with the same $| \mathbf{G} |$: $(110)_R$, $(\bar{2}10)_R$,  $(1\bar{2}0)_R$, and their opposites. For all of these peaks, the phase factor $e^{i \mathbf{G} \cdot \mathbf{\Delta}_l} = 1$. Therefore, for any mixture of R-type stacking, there is a peak at $d=3.433$ \AA\ (given $a = 6.866$ \AA.)

Next, we introduce M-type stacking to the supercell, without changing the spin direction. Setting $\alpha = 1$ for simplicity, the in-plane stacking displacements are 
$-\frac{1}{3} \mathbf{a}_1$, $-\frac{1}{3} \mathbf{a}_2$, and $\frac{1}{3} (\mathbf{a}_1 + \mathbf{a}_2)$. If the additional phase factor from any of these M-type displacements at some location $(11L)_R$ vanishes, then we have constructive interference and a peak will be visible. These additional phase factors are $e^{-i 2 \pi / 3}$ at $(110)_R$, $(\bar{2}10)_R$, and $(1\bar{2}0)_R$. At $L=1$, translation along the $c$-axis will introduce an additional phase factor of $e^{i 2 \pi / 3}$ per layer, resulting in a peak near $d=3.383$ \AA. 

If we, instead, have spin flips across M-type stacking boundaries, an additional phase of $\pm \pi$ is introduced, changing the condition for constructive interference. The additional phase factor for each M-type-stacked layer would be $e^{i \pi/3}$ at $(110)_R$, $(\bar{2}10)_R$, and $(1\bar{2}0)_R$, resulting in constructive interference at $L=-\frac{1}{2}$ and a peak at $d=3.421$ \AA, very close to $(110)_R$ at $d=3.433$ \AA. Thus, any mixture of M-type or R-type stacking would result in a peak of the same intensity within the range of $3.421 \leq d \leq$ 3.433 \AA, leading to the appearance of a single peak given our resolution.

\subsection{Comparison of integrated simulated intensity with data}

\begin{figure}[h]
\begin{center}
\includegraphics[width=8.6cm]
{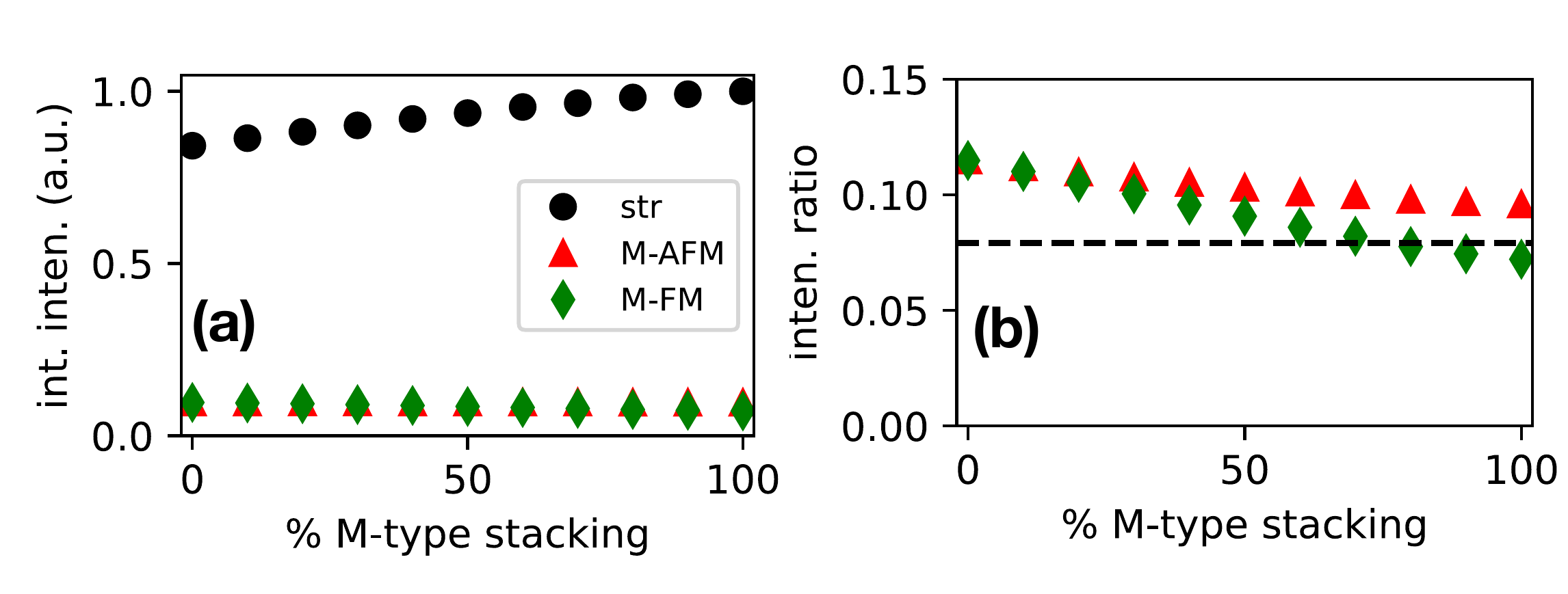}
\end{center}
\caption{(a) Simulated intensities integrated within certain ranges as a function of the percentage of M-type stacking boundaries. The nuclear intensity (``str'') was integrated within $2.5 \leq d \leq 3.52$ \AA, and magnetic intensities for the M-AFM and M-FM models were integrated within $3.36 \leq d \leq 3.52$ \AA\ to capture the $d=3.4$ \AA\ peak. (b) Ratios of the M-AFM and M-FM magnetic intensities to the structural intensities plotted in (a). The dashed line shows the corresponding ratio for the 5 K elastic data.}
\label{fig:S1}
\end{figure}

Here, as a sanity check, we show that the ratio of the magnetic to nuclear intensity in our simulations is consistent with that of our elastic data. In Fig.\ \ref{fig:S1}(a), for a range of percentages of M-type stacking, we plot the integrated intensity of the simulated data for the nuclear intensity integrated within $2.5 \leq d \leq 3.52$ \AA, and the magnetic intensity within $3.36 \leq d \leq 3.52$ \AA\ (i.e., at the $d=3.4$ \AA\ peak) for the M-AFM and M-FM models. In Fig.\ \ref{fig:S1}(b), the ratio of the magnetic $d=3.4$ \AA\ peak intensities to the nuclear intensity is plotted. The two models have similar ratios, with the divergence being due to the larger shift in the $d=3.4$ \AA\ peak position under the M-FM model. The dashed line in Fig.\ \ref{fig:S1}(b) shows the value obtained from the elastic data, in which the magnetic intensity near $d=3.4$ \AA\ (obtained by first subtracting the 70 K data from the 5 K data) is divided by the total intensity at 70 K within $2.5 \leq d \leq 3.52$ \AA. The agreement is clear, though, of course, the measured value is expected to be lower than the ideal value of 3$\mu_B$ per Cr$^{3+}$ ion. Thus, the observed magnetic peak near $d=3.4$ \AA\ is consistent with the amount of nuclear intensity present within $2.5 \leq d \leq 3.52$ \AA, even if this intensity is spread out due to substantial stacking disorder. 

\subsection{$a$-axis lattice constant}

\begin{figure}[h]
\begin{center}
\includegraphics[width=8.6cm]
{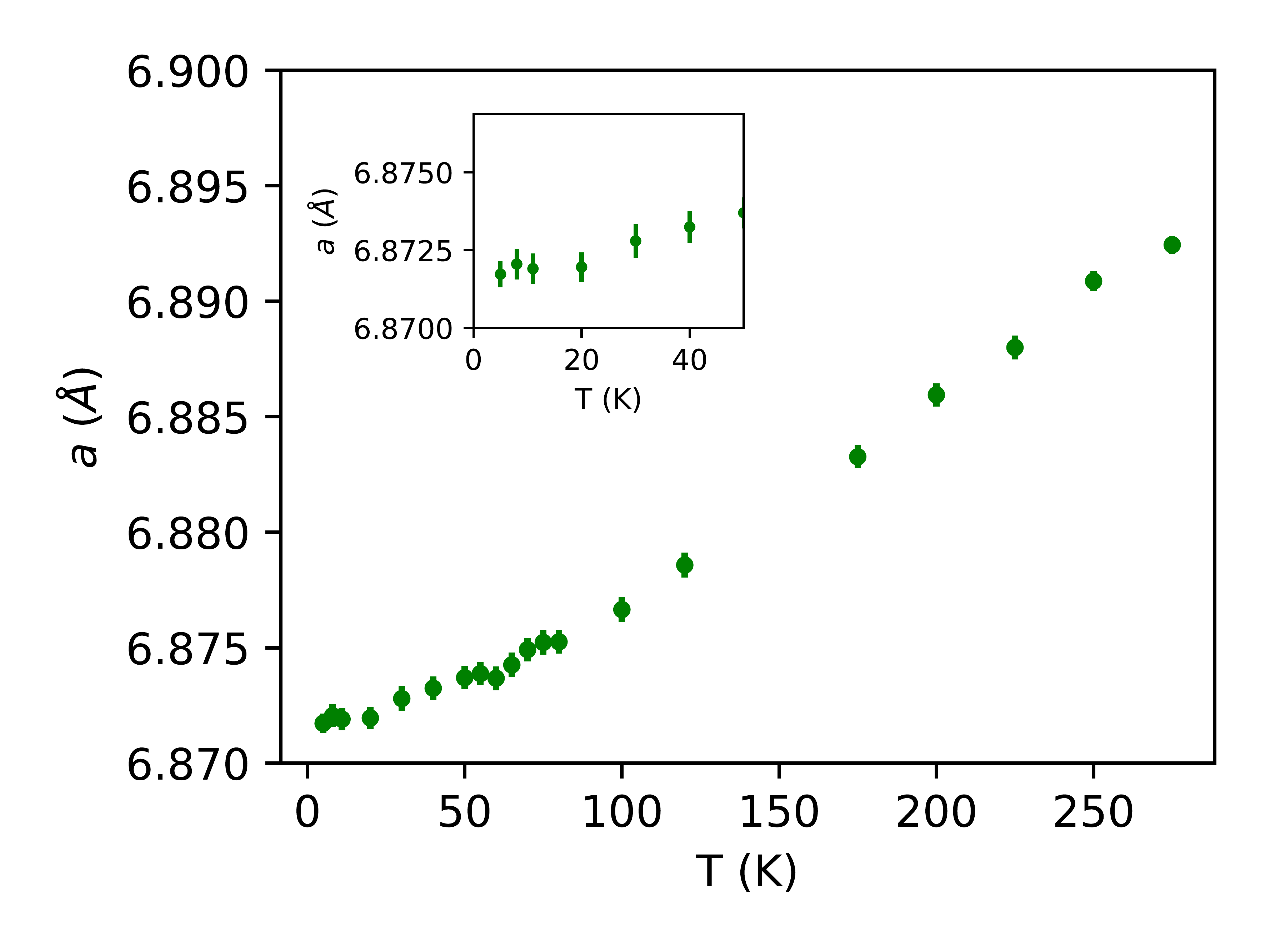}
\end{center}
\caption{The $a$-axis lattice constant  determined from the position of $(300)_R$ in the elastic neutron scattering data. Inset shows the data enlarged.}
\label{fig:aLatticeConstant}
\end{figure}

CrCl$_3$ \cite{schneelochGaplessDiracMagnons2022} and CrBr$_3$ \cite{kozlenkoSpininducedNegativeThermal2021} both exhibit an anomalous in-plane negative thermal expansion (NTE) at low temperature, around $T_C=37$ K for FM CrBr$_3$ but several times $T_N=14$ K for AFM CrCl$_3$. 
In contrast, for CrI$_3$, we show in Figure \ref{fig:aLatticeConstant} that the $a$-axis lattice constant has no clear negative thermal expansion behavior. Surprisingly, this finding is entirely consistent with the same theoretical calculations that predict the in-plane NTE in CrCl$_3$ and CrBr$_3$ \cite{liuNegativeThermalExpansion2022}. 
While our data are rough, there is agreement below 50 K with those calculations, as one can see by comparing Fig.\ 3 in Ref.\ \cite{liuNegativeThermalExpansion2022} with the inset to Fig.\ \ref{fig:aLatticeConstant} that is scaled similarly. 

(For our analysis, we presume that the $a$-axis lattice constant is the same for either R- or M-type stacking, i.e., $a = a_M = a_R$. We determined $a$ from the position of $(300)_R$/$(33\bar{1})_M$/$(060)_M$ in our elastic data. As discussed in the Supplemental Materials of Ref.\ \cite{schneelochGaplessDiracMagnons2022}, if we assume no stacking dependence in the in-plane lattice constants and that $b_M = \sqrt{3} a_M$ (as for a honeycomb lattice with perfect trigonal symmetry), the positions of $(300)_R$ and $(060)_M$ should be the same and solely dependent on $a$. The $d$-spacing position of $(33\bar{1})_M$, meanwhile, deviates slightly from those peaks due to M-type stacking having an in-plane displacement magnitude of $\sim 0.98 a/3$ rather than $a/3$, but this results in a relative error in the $d$-spacing of only $\sim$1.8$\cdot 10^{-5}$, so the discrepancy would be negligible.)

\subsection{Inelastic intensity compared to models}
\begin{figure}[h]
\begin{center}
\includegraphics[width=8.6cm]
{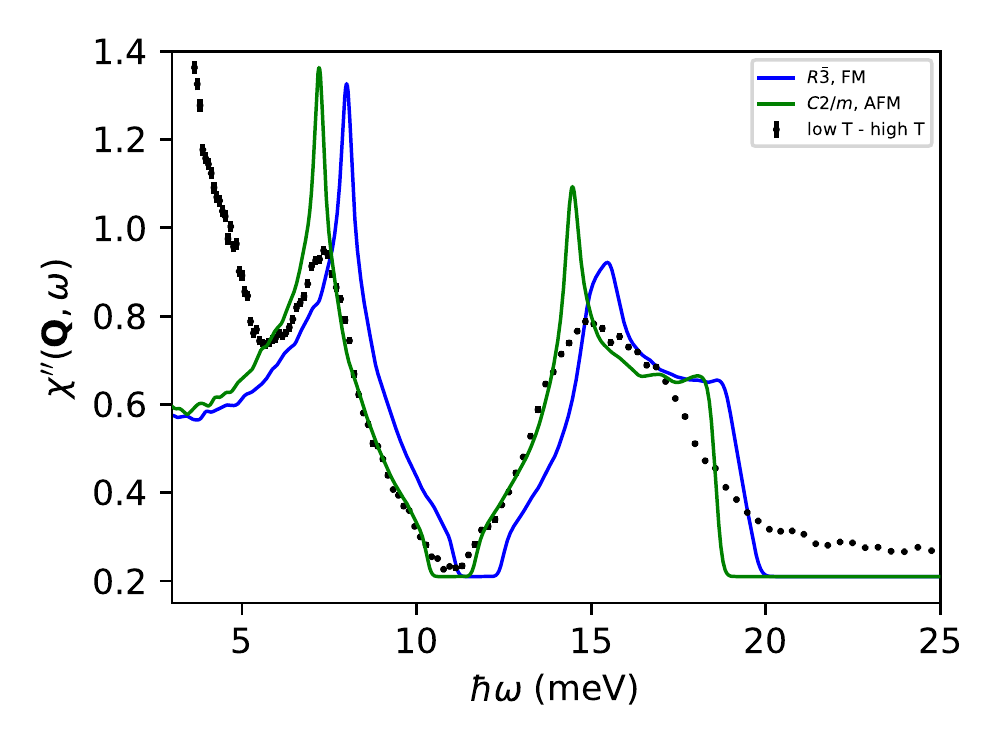}
\end{center}
\caption{Inelastic data as shown in \ref{fig:5}(d), along with $R\bar{3}$ and $C2/m$ models, except that the $C2/m$ model is unshifted in energy.}
\label{fig:inelasticSup}
\end{figure}

In Fig.\ \ref{fig:inelasticSup}, we show the data shown in Fig.\ \ref{fig:5}(d) along with calculations from the $R\bar{3}$ and $C2/m$ models, without shifting the $C2/m$ curve in energy. As discussed in the main text, the only difference between these two models is that the $R\bar{3}$ interlayer magnetic coupling (which sums to -0.59 meV) is replaced with a summed coupling of +0.073 meV, and that the spins are aligned antiferromagnetically in the $C2/m$ model. As we see in Fig.\ \ref{fig:inelasticSup}, as the magnitude of the interlayer magnetic coupling increases, the primary effect is to shift the intensity curve to the right. 

The fact that our data in Fig.\ \ref{fig:5}(d) lines up almost exactly with our calculated $C2/m$ AFM-model curve would seem to imply that our sample is entirely M-type stacking, but we should remember that there is uncertainty in the intraplane interactions of the models that we have borrowed from (as is evident by the changes in refined exchange parameters in subsequent studies as better data were obtained \cite{chenTopologicalSpinExcitations2018,chenMagneticAnisotropyFerromagnetic2020,chenMagneticFieldEffect2021}.)

\subsection{Additional magnetization data}

\begin{figure}[h]
\begin{center}
\includegraphics[width=8.6cm]
{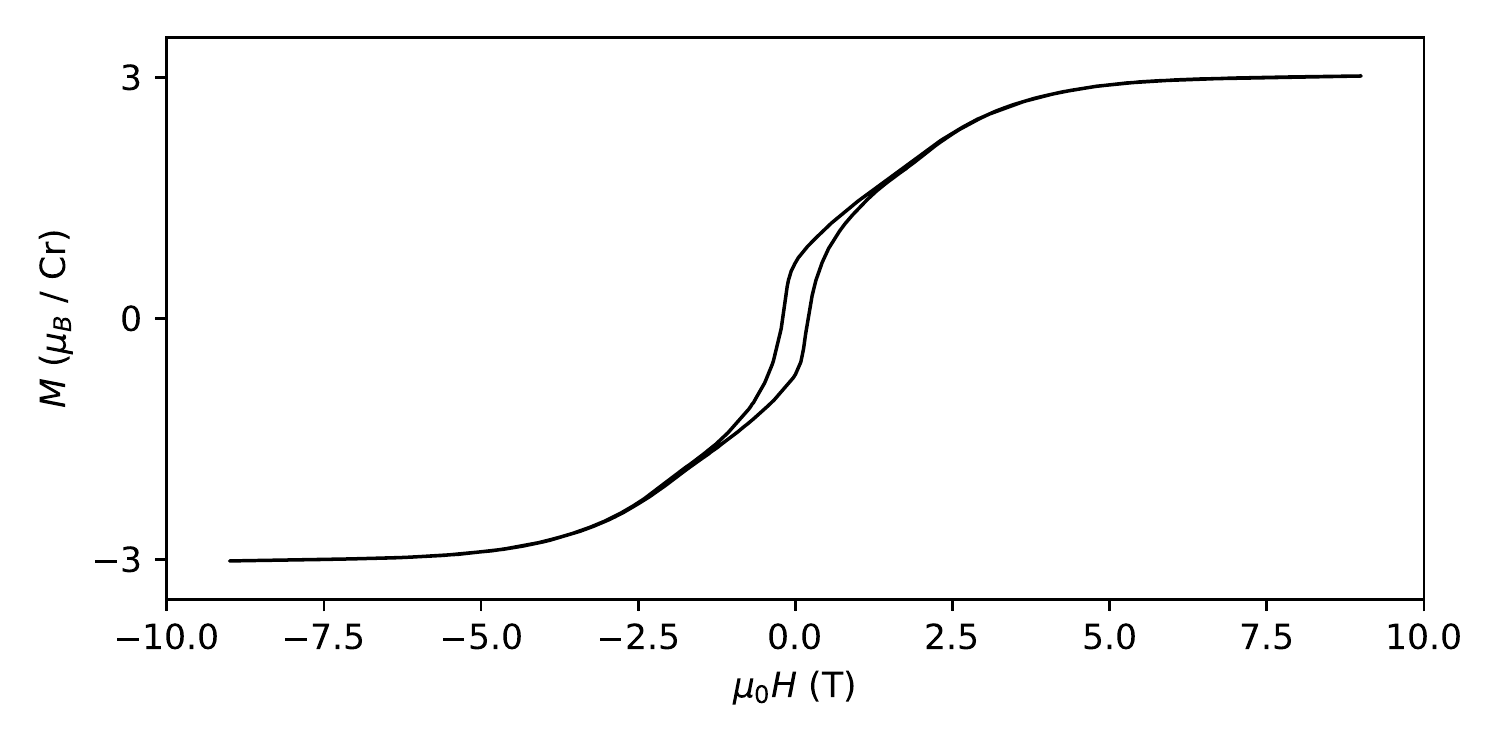}
\end{center}
\caption{Magnetization loop up to $\pm$9 T for a piece of pelletized CrI$_{3}$ powder. Data were collected at 3 K with the same sample and orientation as for the data shown in Fig.\ \ref{fig:5} in the main text.}
\label{fig:9T}
\end{figure}

Here, we show additional magnetization data, taken on the same sample in the same orientation as for Fig.\ \ref{fig:5}, but extending the magnetization loop to $\mu_0 H = \pm$9 T. We see that the saturation magnetization is very close to the expected value of $|M| = 3 \mu_B$ for $S=3/2$ and $g=2$. 

Magnetization vs.\ temperature data at $\mu_0 H = 9$ T (not shown) were also taken; a Curie-Weiss fit to the susceptibility $\chi$ within $250 \leq T \leq 300$ K resulted in a Curie-Weiss temperature of 85.6(1.5) K and a Curie constant of 1.940(10) cm$^3$/(mol K) in CGS units, close to the ideal $S=3/2$, $g=2$ value of 1.875 cm$^3$/(mol K), though outside of the estimated uncertainty.

\end{document}